\documentclass[a4paper,11pt]{elsarticle_WP}

\usepackage{hyperref}
\usepackage{graphicx}
\usepackage{amssymb,amsmath}
\usepackage{bm}
\usepackage[margin=1in]{geometry}
\usepackage{color}
\usepackage{setspace}
\usepackage{multirow}
\usepackage{booktabs}
\usepackage{array}
\usepackage[normalem]{ulem}

\newcolumntype{x}[1]{>{\centering\arraybackslash\hspace{0pt}}p{#1}}

\journal{}

\newcommand{\oneb} {{\bf 1}}

\newcommand{\thetab} {{\boldsymbol{\theta}}}

\newcommand{\nub} {{\boldsymbol{\nub}}}

\newcommand{\varthetab} {{\boldsymbol{\vartheta}}}

\newcommand{\intd} {\textrm{d}}

\newcommand{\zetab} {\boldsymbol{\zeta}}

\newcommand{\Sigmamat} {{\bm \Sigma}}
\newcommand{\Sigmamatt} {\widetilde{\boldsymbol{\Sigma}}}
\newcommand{\Qmatt} {\widetilde{\textbf{Q}}}
\newcommand{\muvect} {\widetilde{\boldsymbol{\mu}}}
\newcommand{\Psib} {{\bm \Psi}}

\newcommand{\Amat} {\textbf{A}}
\newcommand{\Bmat} {\textbf{B}}
\newcommand{\Dmat} {\textbf{D}}

\newcommand{\Qmat} {\textbf{Q}}
\newcommand{\Rmat} {\textbf{R}}

\newcommand{\Cmat} {\mathbf{C}} 
\newcommand{\Jmat} {\mathbf{J}}

\newcommand{\Zmat} {\textbf{Z}}
\newcommand{\Xmat} {\textbf{X}}
\newcommand{\Xvec} {\mathbf{X}}
\newcommand{\Imat} {\textbf{I}}

\newcommand{\Hmat} {\textbf{H}}

\newcommand{\evec} {\textbf{e}}

\newcommand{\vvec} {\textbf{v}}
\newcommand{\svec} {\textbf{s}}
\newcommand{\uvec} {\textbf{u}}

\newcommand{\ff} {\textit{ff}}
\newcommand{\fm} {\textit{fm}}
\newcommand{\mf} {\textit{mf}}
\newcommand{\inv} {\textit{inv}}

\renewcommand{\v}{\mathbf{v}}
\renewcommand{\u}{\mathbf{u}}
\newcommand{\w}{\mathbf{w}}

\newcommand{\x}{\mathbf{x}}

\newcommand{\Yvec}{\mathbf{Y}}

\newcommand{\Zvec}{\mathbf{Z}}
\newcommand{\epsilonb}{\boldsymbol{\varepsilon}}

\newcommand{\E}{\mathrm{E}}
\newcommand{\cov}{\mathrm{cov}}

\newcommand{\Dist}{\mathrm{Dist}}

\newcommand{\diag}{\mathrm{diag}}

\newcommand{\trace}{\mathrm{tr}}

\newcommand{\s}{\mathbf{s}}

\DeclareMathOperator*{\argmax}{arg\,max}

\let\originalleft\left
\let\originalright\right
\renewcommand{\left}{\mathopen{}\mathclose\bgroup\originalleft}
\renewcommand{\right}{\aftergroup\egroup\originalright}

\begin{document}

\begin{frontmatter}

\title{Spatio-temporal bivariate statistical models for atmospheric trace-gas inversion}


\author[Wollongong]{Andrew Zammit-Mangion\corref{correspondingauthor}}
\cortext[correspondingauthor]{Corresponding author}
\ead{azm@uow.edu.au}
\author[Wollongong]{Noel Cressie}
\author[Bristol_Geog]{Anita L. Ganesan}
\author[Bristol_Chem]{Simon O' Doherty}
\author[MetOffice]{Alistair J. Manning}

\address[Wollongong]{National Institute for Applied Statistics Research Australia~(NIASRA), School of Mathematics and Applied Statistics (SMAS), University of Wollongong, Northfields Avenue, Wollongong, NSW 2522, Australia}
\address[Bristol_Geog]{School of Geographical Sciences, University of Bristol, University Road, Bristol, BS8 1SS, UK}
\address[Bristol_Chem]{School of Chemistry, University of Bristol, Cantock's Close, Bristol, BS8 1TS, UK}
\address[MetOffice]{Hadley Centre, Met Office, Fitzroy Road, Exeter, Devon EX1 3PB, UK}

\begin{abstract}
Atmospheric trace-gas inversion refers to any technique used to predict spatial and temporal fluxes using mole-fraction measurements and atmospheric simulations obtained from computer models. Studies to date are most often of a data-assimilation flavour, which implicitly consider univariate statistical models with the flux as the variate of interest. This univariate approach typically assumes that the flux field is either a spatially correlated Gaussian process or a spatially uncorrelated non-Gaussian process with prior expectation fixed using flux inventories (e.g., NAEI or EDGAR in Europe). Here, we extend this approach in three ways. First, we develop a bivariate model for the mole-fraction field and the flux field. The bivariate approach allows optimal prediction of both the flux field and the mole-fraction field, and it leads to significant computational savings over the univariate approach. Second, we employ a lognormal spatial process for the flux field that captures both the lognormal characteristics of the flux field (when appropriate) and its spatial dependence. Third, we propose a new, geostatistical approach to incorporate the flux inventories in our updates, such that the posterior spatial distribution of the flux field is predominantly data-driven. The approach is illustrated on a case study of methane (CH$_4$) emissions in the United Kingdom and Ireland.      
\end{abstract}


\end{frontmatter}


\section{Introduction}\label{sec:Intro}

Atmospheric trace-gas inversion refers to any technique used to predict spatial and temporal fluxes of a gas from observations of mole fractions. Since mole-fraction measurements are affected by weather patterns that are time varying, there is no straightforward relationship between the observations and the fluxes. The relationship, termed a `source-receptor relationship' (SRR), is typically obtained by simulating from a computer model, such as a Lagrangian Particle Dispersion Model (LPDM), which maps the fluxes to the mole fractions \cite{Jones_2007}. The SRR can be characterised through a bivariate function $b_t(\svec,\uvec)$ where the spatial locations $\svec,\u \in \mathbb{R}^2$ are in a given domain of interest, $D \subset \mathbb{R}^2$, and $t \in \{1,2,\dots\}$ is a discrete-time index. 

\begin{figure}[!ht]
\begin{center}
\includegraphics[width=3in]{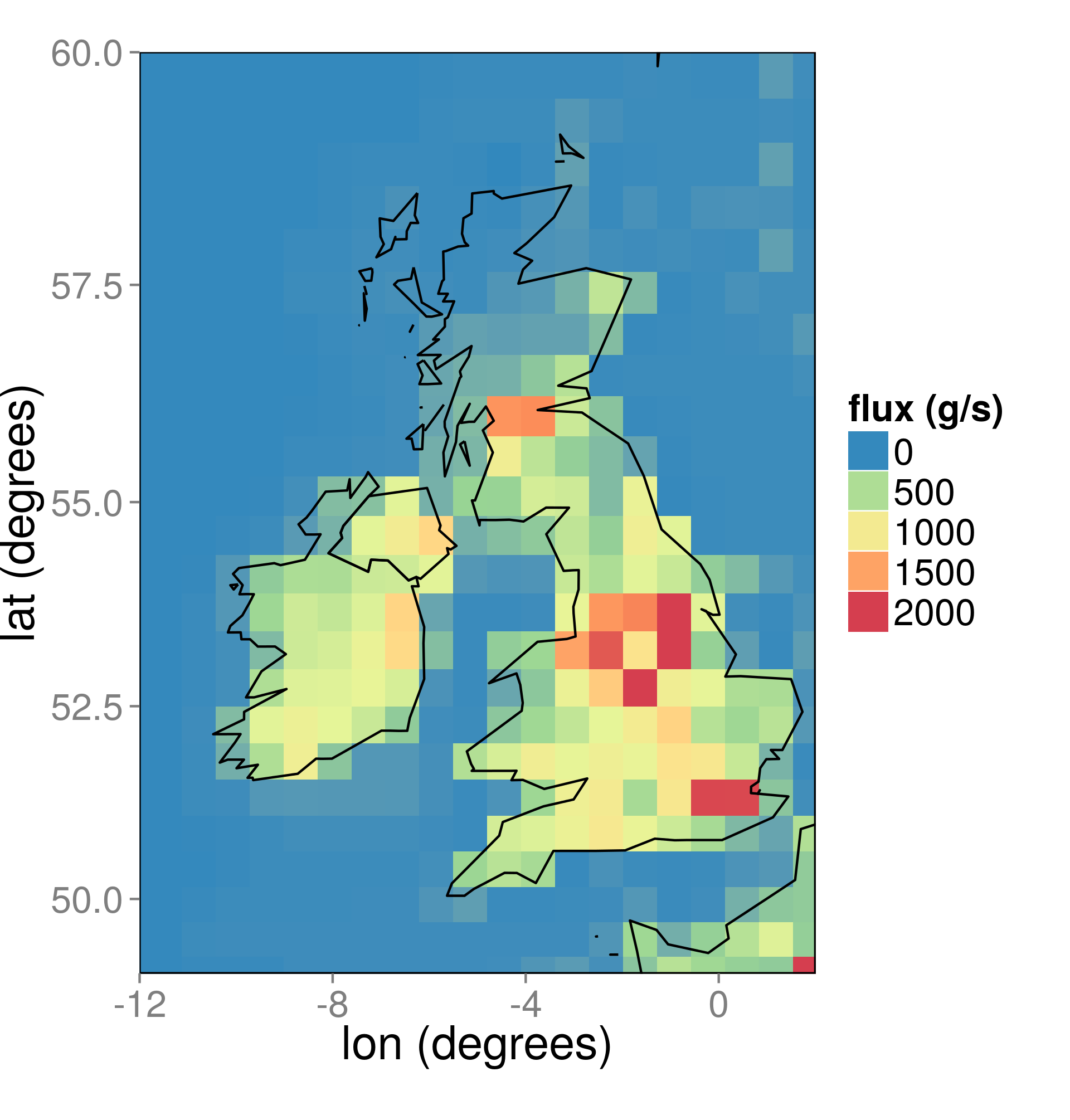}  
\includegraphics[width=3in]{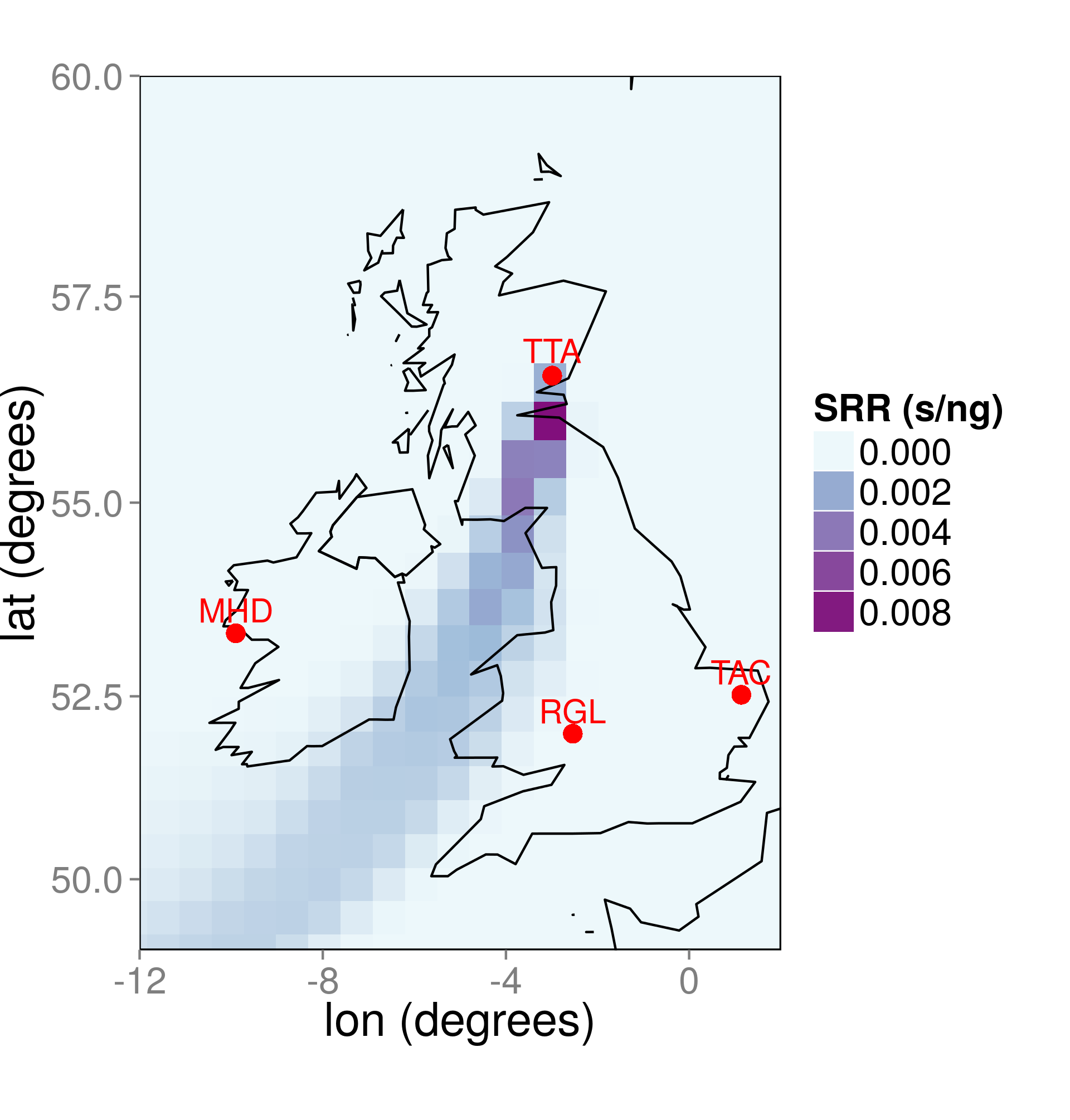}
\includegraphics[width=5.5in]{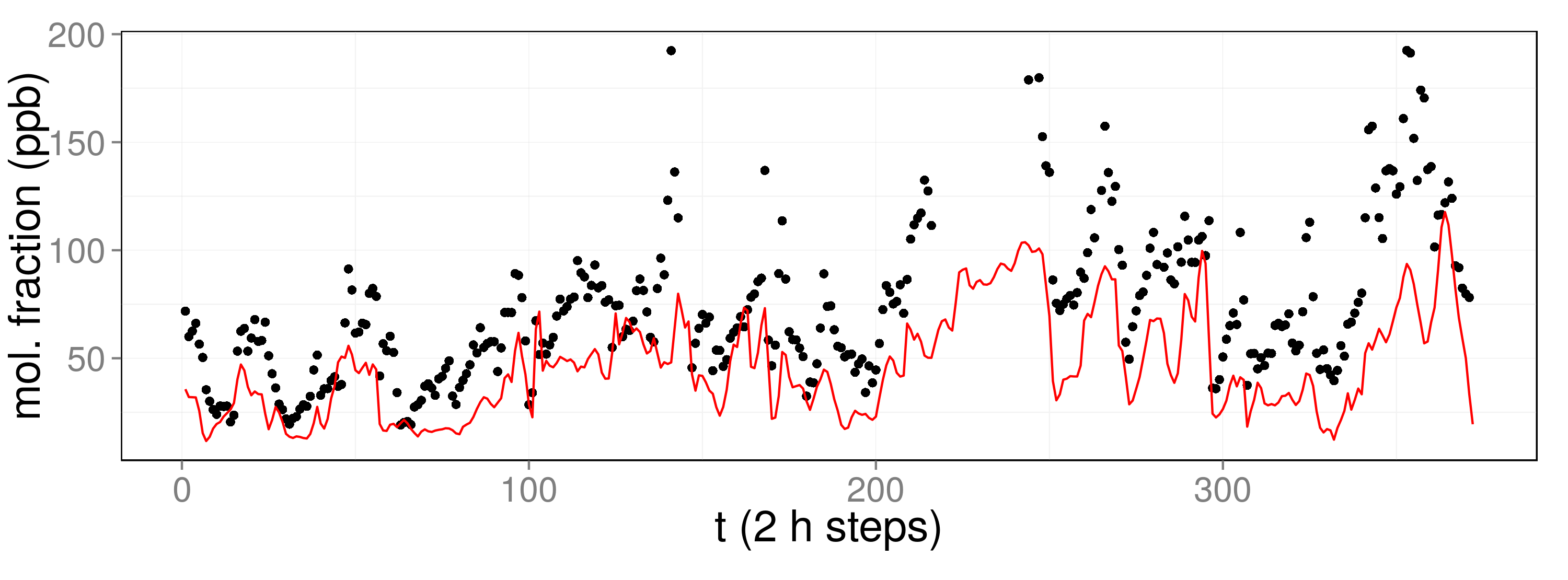}  
	\caption{Top-left panel: Total flux by grid cell in g s$^{-1}$ obtained by combining methane inventories (see \cite{Ganesan_2015} for details) for January 2014. Top-right panel: The source-receptor relationship on 01 January 2014 at 00:00, $b_1(\s_{TTA},\cdot)$, obtained from the UK Met Office's Lagrangian Particle Dispersion Model (LPDM), Numerical Atmospheric-dispersion Modelling Environment (NAME), where $\s_{TTA}$ is the coordinate vector of the Angus measurement station (\textsc{TTA}), in Scotland. Also shown are the three other stations used in this study, Mace Head (\textsc{MHD}), Ridge Hill (\textsc{RGL}), and Tacolneston (\textsc{TAC}). Bottom panel: Measurements of methane mole fraction in parts per billion (ppb) at \textsc{TAC} for January 2014, following background-removal (black dots), together with a straightforward prediction (red line) using NAME and the methane flux inventory. Each time step corresponds to an interval of 2 h.} \label{fig:intro}
\end{center}
\end{figure}

Figure \ref{fig:intro}, top-left panel, shows the UK and Ireland methane (CH$_4$) total fluxes by grid cell, in units g s$^{-1}$, obtained from a combination of flux inventories. Here, the component inventories are principally the National Atmospheric Emissions Inventory (NAEI) \cite{NAEI} and the Emissions Database for Global Atmospheric Research version 4.2 (EDGAR) \cite{Edgar}; see \cite{Ganesan_2015} for details. Figure \ref{fig:intro}, top-right panel, shows an evaluation of $b_t(\svec,\uvec)$, in units s ng$^{-1}$,  of methane fluxes, for $\svec$ located at a mole-fraction monitoring station at Angus, Scotland (TTA), where $\uvec$ takes values on a discrete lattice, and where $t=1$ (01 January 2014 at midnight). Figure \ref{fig:intro}, bottom panel, shows the two-hourly averaged observations in parts per billion (ppb) at TTA, following background-removal (see Section \ref{sec:uk} for details), for all of January 2014. In atmospheric trace-gas inversion, the aim is to recover a flux map, such as that in Figure \ref{fig:intro}, top-left panel, from time-series observations taken at various monitoring stations and from the collection of atmospheric simulations (one such is shown in Figure \ref{fig:intro}, top-right panel) that establish the SRR.

Atmospheric trace-gas inversion is an ill-posed problem. Consequently, ``small uncertainties in the observational data correspond to much higher uncertainty in the emission[s]'' \cite{Michalak_2004}. However, in addition to observation and flux-field uncertainties, there is a third source of uncertainty arising from the use of an atmospheric model that does not perfectly match the true SRR due to physical parameterisations, solver discretisations, etc.  Sometimes this third term is called the \emph{discrepancy term}, and failure to acknowledge it can lead to over-confident predictions \cite{Brynjarsdottir_2014,Kennedy_2001}. Until very recently, this discrepancy was not considered separately from the observation error; see \citep{Ganesan_2015}. However, it is crucial to distinguish between the errors due to observations and those due to model misspecification, as these are likely to have different statistical properties. 



A critical contribution of our work is to formalise the insight in \cite{Ganesan_2015} and treat the mole-fraction field as a second variable of interest. The consideration of modelling a mole-fraction field in addition to the flux field through a discrepancy term results in a \emph{bivariate} model (see \cite{Genton_2015} for a recent review on such models). The bivariate model brings two advantages to this problem of trace-gas inversion: The locations at which the mole-fraction field is modelled need not coincide with the data points. This in turn allows the predictive distribution of mole fraction at any unobserved locations to be found, and then averaged over any subset of the spatio-temporal domain, with relative ease (provided the SRR is available for these locations/domain). The other advantage is that the decoupling leads to computationally efficient methods in spatial statistics that can be used to scale up the inversion to large, remote-sensing datasets.


Another contribution of our work is to introduce the lognormal spatial process as a prior distribution for the flux field. This model acts as a bridge between the two sides of the dichotomous literature that either assumes spatial (possibly truncated) Gaussian-process priors (e.g., \cite{Miller_2013}) or spatially uncorrelated lognormal priors (e.g., \cite{Ganesan_2014}). A lognormal spatial process is attractive, as it is able to capture both (i) the nonnegativity and heavy tails in the distribution of the flux (valid for some trace gases such as methane) and (ii) the spatial correlation of the flux field. We show that expectations and covariances for both the flux field and the mole-fraction field can be obtained analytically if the flux field is defined as a lognormal process, by directly applying results from the univariate case \citep[][p.~135]{Cressie_2006,Cressie_1993}.

The third contribution of our work is to propose a new way to carry out assimilation in atmospheric trace-gas inversion. Frequently, the prior expectation of the flux process is set from one or more inventories that are many times unreliable and that may have unquantifiable effects on a posterior assessment. Here, we only use \emph{characteristics} of the inventories, namely the spatial length scales and the marginal variance, while setting the prior expectation to be spatially constant. In this way, the inventory fluxes are not used directly in the assimilation, and the spatial distribution of fluxes obtained from the posterior expectation will be predominantly data-driven. Our contribution addresses a concern in \cite[][Section 5.2]{Miller_2014} that suggests that such an approach is difficult when prior distributions are non Gaussian.


The article is structured as follows. In Section \ref{sec:model}, we discuss the three contributions outlined above. In Section \ref{sec:inference}, we detail our approach to inferring the fields of interest using a combination of approximate inferential methods. The proposed framework is then assessed in Section \ref{sec:results}, first in a study in one-dimensional space with simulated datasets, and then for emissions prediction in the UK and Ireland using the four measurement stations illustrated in Figure \ref{fig:intro}, top-right panel. Section \ref{sec:discussion} contains conclusions and an outline of future research directions.

\section{Theory}\label{sec:model}

The notation we use corresponds to that commonly found in the spatial-statistics literature (e.g., \cite{Cressie_2011}). Here, stochastic processes are denoted using regular typeface, while bold typeface is used to denote vectors. The symbol $Z$ denotes observations while $Y$ denotes processes (fields). The notation $Y(\cdot)$ is shorthand for `the process $Y$ over the given spatial domain' while $Y(\svec)$ is the process $Y$ evaluated at a specific location $\svec$. The subscripts $m$ and $f$ are used for fields or domains associated with the mole-fraction field and the flux field, respectively, while the subscript $t$ denotes a discrete-time index. Therefore $Y_f(\cdot)$ denotes the flux field, while $Y_{m,t}(\cdot)$ denotes the spatio-temporal mole-fraction field at time-index $t$. Spatial domains are notated using the letter $D$; the superscript $O$ is used to denote observation domains and the superscript $L$ is used to denote process (discretised onto a lattice) domains. The cardinality of a set is denoted using $|\,\cdot\,|$, while $[A|B]$ denotes the conditional probability distribution of $A$ given $B$. 

For simplicity, as in \cite{Ganesan_2015,Ganesan_2014}, we will assume throughout that the flux is a \emph{spatial} process over a short time period; the development of models for \emph{spatio-temporal} flux processes will be considered elsewhere. Recall that we use the subscript $f$ to denote `flux,' however because in this paper we constrain the fluxes to be positive \emph{a priori}, we use the words `flux' and `emissions' interchangeably.

\subsection{Bivariate modelling: Treating mole fraction as a second variate}\label{sec:bivariate}

In this section we discuss the univariate- and bivariate-modelling approaches to atmospheric trace-gas inversion, highlighting the key differences between the two. This discussion is used to motivate the advantages of the bivariate model, namely (i) interpretability, (ii) the possibility to compute the predictive distribution of mole fraction anywhere in, or averaged over any subset of,  the domain of interest, and (iii) computational benefits arising from the decoupling.

Denote the mole-fraction observation at location $\s$ and time $t$ as $Z_{m,t}(\svec)$, and define $D_{m,t}^O$ as the set of measurement locations at time $t$. Further, denote the collection of all observation locations as $D^O_m \equiv \bigcup_{t \in \mathcal{T}} D^O_{m,t}$, where $\mathcal{T} \equiv \{1,\dots,T\}$ is the temporal domain and $T > 1$ is the number of time steps. There are typically very few or no flux observations; flux prediction usually proceeds by using the following \emph{univariate model} (or its discretised equivalent) for inversion \cite{Miller_2013,Rigby_2011, Ganesan_2014,Stohl_2009,Thompson_2014}:
\begin{equation}\label{eq:uv}
Z_{m,t}(\svec) = \int_D b_t(\svec,\uvec)Y_f(\uvec)\intd\u + e_{m,t}(\svec); \quad \svec \in D^O_{m,t}\subset D;\quad t \in \mathcal{T},
\end{equation}
where $D$ is the spatial domain, $e_{m,t}(\svec)$ is (Gaussian) observation error, $b_t(\svec,\uvec)$ is the interaction function, or SRR, that returns quantities with dimensions [time][mass]$^{-1}$, and $Y_f(\cdot)$ is the flux field  that returns quantities with dimensions [mass][time]$^{-1}$[area]$^{-1}$. The field $Y_f(\cdot)$ is a stochastic process that, as we discuss below in Section \ref{sec:lognormal}, has statistical properties that reflect those of the underlying process (e.g., positivity).  Note that $\{Y_f(\uvec): \uvec \in D\}$ is the flux density, and it is different from the flux by grid cell $\{ A(\uvec) Y_f(\uvec): \uvec \in D^L\}$ (where $A(\uvec)$ is the grid cell area at $\uvec$ and $D^L$ is some discrete subset of $D$) that returns quantities with dimensions [mass][time]$^{-1}$. Note also that there is no mention of the complete mole-fraction field in \eqref{eq:uv}, only observations of the mole-fraction field at specific locations. 

In our \emph{bivariate model}, the object of interest shifts from the \emph{observations} of mole fraction to the mole-fraction \emph{field} itself, which we model as 
\begin{equation}\label{eq:mv}
Y_{m,t}(\svec) = \int_D b_t(\svec,\uvec)Y_f(\uvec)\intd\u + \zeta_{t}(\svec); \quad \svec \in D^L_m;\quad t \in \mathcal{T},
\end{equation}
\noindent where $Y_{m,t}(\cdot)$ is the mole-fraction field and $D^L_m$ is a user-defined set of discrete locations in $D$ at which $Y_{m,t}$ is evaluated.  The additive component, $\zeta_{t}(\svec)$, captures the spatio-temporal discrepancy (due to mole-fraction background) and variation in the true mole-fraction field that cannot be explained by the flux field through the SRR. Note that it is different from $e_{m,t}(\svec)$ in \eqref{eq:uv}, which represents all the unexplained signal, since it explicitly accounts for \emph{model} misspecification. Finally, the observations available for the inversion are not the field itself, but a noisy, incomplete version of it:
\begin{equation}\label{eq:obs}
Z_{m,t}(\svec) = Y_{m,t}(\svec) + \varepsilon_{m,t}(\svec); \quad \svec \in D^O_{m,t},
\end{equation}
\noindent where $\varepsilon_{m,t}(\cdot)$ is a measurement-error process.



There are three important differences between \eqref{eq:uv} and \eqref{eq:mv}. First, \eqref{eq:mv} is defined on $D^L_m$, while \eqref{eq:uv} is only defined for $D_{m,t}^O \subset D$; that is, the locations at which mole fraction is inferred need not (but could) coincide with the observation locations. Second, due to the decoupling, the locations $\svec \in D^L_m$ at which the SRR is evaluated are not dependent on the observation locations; the case of $|D^L_m| \ll |D^O_m|$ occurs for regional studies that use computationally-intensive LPDMs to simulate the SRR.
Third, the mole-fraction observations do not appear in \eqref{eq:mv}. This is computationally beneficial as it is usually reasonable to assume that observations are conditionally independent given the mole-fraction field, but not necessarily given only the flux field. This conditional independence implies that the conditional covariance matrix of $\{Z_{m,t}(\svec)\}$ given the mole-fraction field is diagonal, resulting in simplified matrix computations.








Consider a discretisation of $D$ for the flux field, denoted $D^L_f$ (that need not be identical to $D^L_m$). Define


\noindent\begin{minipage}{.5\linewidth}
\begin{align*}
\Zvec_{m,t} &\equiv (Z_{m,t}(\svec): \, \svec \in D_{m,t}^O)',\\
\Zvec_{m} &\equiv (\Zvec_{m,t}': \, t \in \mathcal{T})',\\
\Yvec_{m,t} &\equiv (Y_{m,t}(\svec): \, \svec \in D^L_m)', \\
\Yvec_{m} &\equiv (\Yvec_{m,t}': \, t \in \mathcal{T})', \\
\Yvec_{f} &\equiv (Y_{f}(\svec): \, \svec \in D^L_f)',
\end{align*}
\end{minipage}
\noindent\begin{minipage}{.5\linewidth}
\begin{align*}
\evec_{m,t} &\equiv(e_{m,t}(\svec): \, \svec \in D_{m,t}^O)',\\
\evec_{m} &\equiv (\evec_{m,t}': \, t \in \mathcal{T})',\\
\epsilonb_{m,t} &\equiv (\varepsilon_{m,t}(\svec): \, \svec \in D_{m,t}^O)',\\
\epsilonb_{m} &\equiv (\epsilonb_{m,t}': \, t \in \mathcal{T})',\\
\zetab_{t} &\equiv (\zeta_{t}(\svec): \, \svec \in D^L_m)',\\
\zetab &\equiv (\zetab_{t}': \, t \in \mathcal{T})',
\end{align*}
\end{minipage}

\vspace{0.2in}

\noindent where the `prime' symbol is used to denote the transpose operation. Then, in the univariate model \eqref{eq:uv}, the relationship between the mole-fraction \emph{observations} and the flux field is approximated by
\begin{equation}\label{eq:uv2}
\Zvec_{m,t} \simeq \Bmat_{U,t}\Yvec_{f} + \evec_{m,t}; \quad t \in \mathcal{T},
\end{equation}
\noindent where $\Bmat_{U,t} \equiv (A(\uvec) b_t(\svec,\uvec): \svec \in D^O_{m,t}; \uvec \in D^L_f)$ is a $|D_{m,t}^O| \times |D_f^L|$ matrix that approximates the integral in \eqref{eq:uv}, and $\{A(\uvec): \uvec \in D^L_f\}$ are integration weights with dimension, [area]. In the bivariate model, the relationship between the mole-fraction \emph{field} and the flux field, \eqref{eq:mv}, is approximated by
\begin{equation}\label{eq:mv2}
\Yvec_{m,t} \simeq \Bmat_{B,t}\Yvec_{f} + \zetab_{t}; \quad t \in \mathcal{T},
\end{equation}
\noindent where 
\begin{equation}\label{eq:BMt}
\Bmat_{B,t} \equiv (A(\uvec) b_t(\svec,\uvec): \svec \in D^L_m; \uvec \in D^L_f)
\end{equation}
is a $|D_m^L| \times |D_f^L|$ matrix that approximates the integral in \eqref{eq:mv}. The relationship between the mole-fraction observations and the mole-fraction field is
\begin{equation}\label{eq:obs_discrete}
\Zvec_{m,t} = \Cmat_t\Yvec_{m,t} + \epsilonb_{m,t}; \quad t \in \mathcal{T},
\end{equation}
\noindent where $\Cmat_t$ is a $|D_{m,t}^O| \times |D_m^L|$ incidence matrix that indicates where the observation locations $D_{m,t}^O$ are in $D_m^L$.

The first advantage of the bivariate approach is the ability to compute the predictive distribution of the mole-fraction field at locations, or over space-time domains, where the field is not directly observed (note that the SRR still needs to be available for these locations or domains). In the univariate approach, \eqref{eq:uv2}, we are able to compute (through a standard application of Bayes' rule), the probability distribution of the flux field given the mole-fraction observations, that is, $[\Yvec_f \mid \Zvec_m]$. Following this, we could try to obtain the predictive distribution of the mole fractions at time $t$ through $[\Bmat_{U,t}\Yvec_f \mid  \Zvec_m]$, but this does not take into account the discrepancy $\zeta_t(\svec)$. In the bivariate case, with \eqref{eq:mv2} and \eqref{eq:obs_discrete} we can compute the \emph{joint} posterior distribution $[\Yvec_f, \Yvec_m \mid \Zvec_m]$ at locations $D_m^L \times D_f^L$, from which we can in principle obtain both the posterior distribution of the flux field $[\Yvec_f\mid \Zvec_m]$ \emph{and} that of the mole-fraction field $[\Yvec_m \mid \Zvec_m]$. We may also easily find the posterior distribution of any affine transformation of the random vector $\Yvec_m$, and hence it is straightforward to carry out inferences over a spatial or temporal aggregation  of the mole-fraction field if desired.


The second advantage of the bivariate approach is computational. In many applications it is reasonable to assume that the observation error $\epsilonb_{m}$ is uncorrelated. However, this assumption does not hold for $\evec_{m}$ (in the univariate case), since $e_{m,t}(\svec)$  should incorporate spatio-temporal correlations arising indirectly from the discrepancy term \cite{Ganesan_2014}. Since the observations $\Zvec_m$ are (in practice) spatio-temporally irregular, there is no straightforward way, without modifying the underlying model (e.g., through covariance tapering \cite{Kaufman_2008}), to define $\cov(\evec_{m,t})$ such that it is sparse or such that it can be decomposed into simpler components. On the other hand, $Y_{m,t}(\cdot)$ (equivalently $\zeta_{t}(\cdot)$) can be discretised in such a way that covariance matrices associated with $\zeta_t(\svec)$ can be easily stored and inverted. For example, in Section \ref{sec:models}, we use a regular space-time grid for the discretisation, and we exploit the properties of the Kronecker product to simplify the computation as in \cite{Ganesan_2015}. 

\subsection{The lognormal bivariate model and its properties}\label{sec:lognormal}


To date, most studies have assumed either that $Y_f(\cdot)$ is a Gaussian process or that $Y_f(\cdot)$ is spatially uncorrelated and non-Gaussian. 
Non-Gaussianity is generally important to model since the spatial distribution of fluxes tends to exhibit a heavy right tail and, for some trace gases such as methane, it is reasonable to further assume that it has only positive support. However, it is easy to see from flux inventories that often the spatially-uncorrelated assumption is unrealistic. In this section we present a new model, namely a lognormal spatial process, that can cater for both non-Gaussianity and for spatial correlations that appear in the flux field. We then embed this in the bivariate model and discuss the model's resulting first-order and second-order properties.


Assume that $Y_f(\cdot)$ is a spatial process such that $\widetilde{Y}_f(\cdot) \equiv \ln Y_f(\cdot)$ is a Gaussian process \cite{Rasmussen_2006}. Then $Y_f(\cdot)$ is termed a \emph{lognormal spatial process}; clearly, $P(Y_f(\svec) \le 0) = 0,$ for $\svec \in D$, so that $Y_f(\cdot)$ is almost surely positive everywhere.  Assume further that $\E(\widetilde{Y}_f(\s)\mid\varthetab) \equiv \widetilde\mu_f(\s\mid \varthetab)$ and $\cov(\widetilde{Y}_f(\s),\widetilde{Y}_f(\u)\mid\varthetab) \equiv \widetilde{C}_{\ff}(\s,\u\mid  \varthetab)$, where $\varthetab$ is a vector of unknown parameters. Then from \cite{Aitchison_1957},
\begin{align}
\mu_f(\s\mid\varthetab) &\equiv \E(Y_f(\s)\mid\varthetab) \equiv  \exp(\widetilde{\mu}_f(\svec\mid \varthetab) + (1/2)\widetilde{C}_{\ff}(\s,\s\mid \varthetab)); \quad \svec \in D, \label{eq:ln1}\\
 C_{\ff}(\s,\u\mid \varthetab) &\equiv \cov(Y_f(\s),Y_f(\u)\mid\varthetab) \equiv \mu_f(\s\mid \varthetab)\mu_f(\u\mid \varthetab)[\exp(\widetilde{C}_{\ff}(\s,\u\mid \varthetab)) - 1] ; ~\svec,\uvec \in D. \label{eq:ln2}
\end{align}
\noindent That is, the mean and covariances of the lognormal spatial process are adjusted versions of the exponentiated mean and covariances of the Gaussian spatial process, and thus they can be easily computed.

We only need to specify two components in \eqref{eq:ln1} and \eqref{eq:ln2}: the prior mean and the prior covariance functions in the `log space,' that is, $\widetilde{\mu}_f(\svec\mid \varthetab)$ and $\widetilde{C}_{\ff}(\svec,\uvec\mid \varthetab);~\svec,\uvec \in D$. The way in which we do so does not differ substantially from when Gaussian spatial processes are used to model the prior and, in this regard, research to date falls into two broad categories. First, in works that employ a data-assimilation framework,  $\widetilde{\mu}_f(\svec\mid \varthetab)$ is typically set from values in a flux inventory, whilst $\widetilde{C}_{\ff}(\svec,\uvec\mid \varthetab);~\svec,\uvec \in D$, is chosen through expert judgment (e.g., \cite{Gurney_2002}). Second, in works that employ a geostatistical framework, $\widetilde{\mu}_f(\svec\mid \varthetab)$ is defined as a linear combination of regressors, whilst $\widetilde{C}_{\ff}(\svec,\uvec\mid \varthetab);~ \svec,\uvec \in D$, is a standard univariate covariance function (e.g., the stationary and isotropic exponential covariance function \cite{Miller_2013,Gourdji_2008}). In Section \ref{sec:data-feature}, we offer an enhanced approach to eliciting the first-order and second-order moments of $Y_f(\cdot)$.

Given \eqref{eq:mv}, we are able to obtain closed-form expressions for the expectation and covariance function of the mole-fraction field at time $t$. These are,
 \begin{align} \label{eq:mu}
\mu_{m,t}(\s\mid \varthetab) &\equiv \E(Y_{m,t}(\svec)\mid \varthetab) = \int_D b_t(\svec,\vvec)\mu_f(\vvec\mid \varthetab)\intd\vvec + \E(\zeta_t(\svec)); \quad \svec \in D, 
\end{align}
where we deliberately allow the possibility that $\E(\zeta_t(\svec)) \ne 0$; and
\begin{align}
C_{mm,t}(\s,\u\mid \thetab,\varthetab) &\equiv \cov(Y_{m,t}(\s),Y_{m,t}(\u) \mid \thetab, \varthetab) \nonumber \\
& = C_{\zeta\zeta,t}(\s,\u\mid \thetab) + \int_D\int_D b_t(\s,\v)C_{\ff}(\v,\w\mid \varthetab)b_t(\u,\w)\intd\v\intd\w; \quad \s,\u \in D,
\end{align}
\noindent where $C_{\zeta\zeta,t}(\s,\u\mid \thetab) \equiv \cov(\zeta_t(\svec),\zeta_t(\uvec) \mid \thetab)$, and $\thetab$ is a vector of unknown parameters appearing in the covariance function of the discrepancy term. Again, from \eqref{eq:mv}, the cross-covariance functions between the mole-fraction field at time $t$ and the flux field are, 
\begin{align} 
C_{\fm,t}(\s,\u\mid \varthetab) &\equiv \cov(Y_f(\s),Y_{m,t}(\u) \mid \varthetab) = \int_D C_{\ff}(\s,\w\mid \varthetab)b_t(\u,\w) \intd\w; \quad \s, \u \in D, \\
C_{\mf,t}(\s,\u\mid \varthetab) &\equiv \cov(Y_{m,t}(\s),Y_f(\u) \mid \varthetab) = \int_D C_{\ff}(\w,\u\mid \varthetab)b_t(\s,\w) \intd\w; \quad \s, \u \in D.  \label{eq:Cmf}
\end{align}

Equations \eqref{eq:mu}--\eqref{eq:Cmf} are identical to those given in \cite{Cressie_2015} with $\mu_f$ and $C_{\ff}$ given by \eqref{eq:ln1} and \eqref{eq:ln2}, respectively. 
Now we can write down the expectation and covariance of the joint model at time $t$ as
\begin{align}\label{eq:joint}
\Yvec_t(\cdot) \sim \Dist\begin{pmatrix} \begin{pmatrix} \mu_f(\cdot) \\ \mu_{m,t}(\cdot) \end{pmatrix},\begin{pmatrix}  C_{\ff}(\cdot,\cdot) &  C_{fm,t}(\cdot,\cdot) \\  C_{mf,t}(\cdot,\cdot) &  C_{mm,t}(\cdot,\cdot) \end{pmatrix} \end{pmatrix},
\end{align}
\noindent where $\Yvec_t(\cdot) \equiv (Y_f(\cdot), Y_{m,t}(\cdot))'$, Dist($\cdot,\cdot$) is a distribution that here is not Gaussian but which features the first two moments, and the terms in \eqref{eq:joint} are given by \eqref{eq:mu}--\eqref{eq:Cmf}. We stress that, unlike in a bivariate Gaussian process, here the joint model $[\Yvec_t(\cdot)]$ given by \eqref{eq:joint} is not fully specified by the mean and covariance functions. 

We have derived above the expectation and covariance of $\Yvec_t(\cdot)$, that is, of the bivariate process at a single time point $t$. In practice, we need to find the expectation and covariance of 
\begin{equation}
\Yvec(\cdot)\equiv (Y_f(\cdot), (Y_{m,t}(\cdot) : t \in \mathcal{T}))'.
\end{equation}
These can be obtained using the same ideas used to construct \eqref{eq:mu}--\eqref{eq:Cmf} and the mole-fraction covariances $\cov(Y_{m,t}(\cdot),Y_{m,t'}(\cdot)); t,t' \in \mathcal{T}$.


\subsection{Inventories for estimating the spatial lognormal characteristics of the flux field}\label{sec:data-feature}

One would ideally estimate from the data all unknown parameters appearing in the model. However, this can be difficult for parameters affecting fields that are not directly observed (in our case $Y_f(\cdot)$) and in ill-posed problems such as that considered here. In other fields of study, this problem has been rectified by estimating the parameters from deterministic models. In \cite{Calder_2011}, for example, parameters appearing in a dynamic stochastic model of aerosol optical depth were estimated from a chemical-tracer numerical model. In \cite{Zammit_2014}, the varying spatial-length scales of precipitation patterns were estimated from a regional climate model. In the present study, we propose estimating the spatial-length scale and the marginal variance of the process $Y_f(\cdot)$ from flux inventories. Importantly, unlike several works in atmospheric trace-gas inversion, we do not fix the prior expectation of the flux process to be equal to values given in an inventory, since it is often found that inventories can be biased due to errors in activity data or emissions factors. Instead, we set $\widetilde\mu_f(\cdot) = c$, where $c \in \mathbb{R}$ is a spatially-invariant constant evaluated below. We now show the procedure for obtaining $c$ and the parameters in $\widetilde{C}_\ff(\cdot,\cdot)$ from the inventory shown in Figure \ref{fig:intro}, top-left panel.

\begin{figure}[!t]
\begin{center}
\includegraphics[width=3.0in]{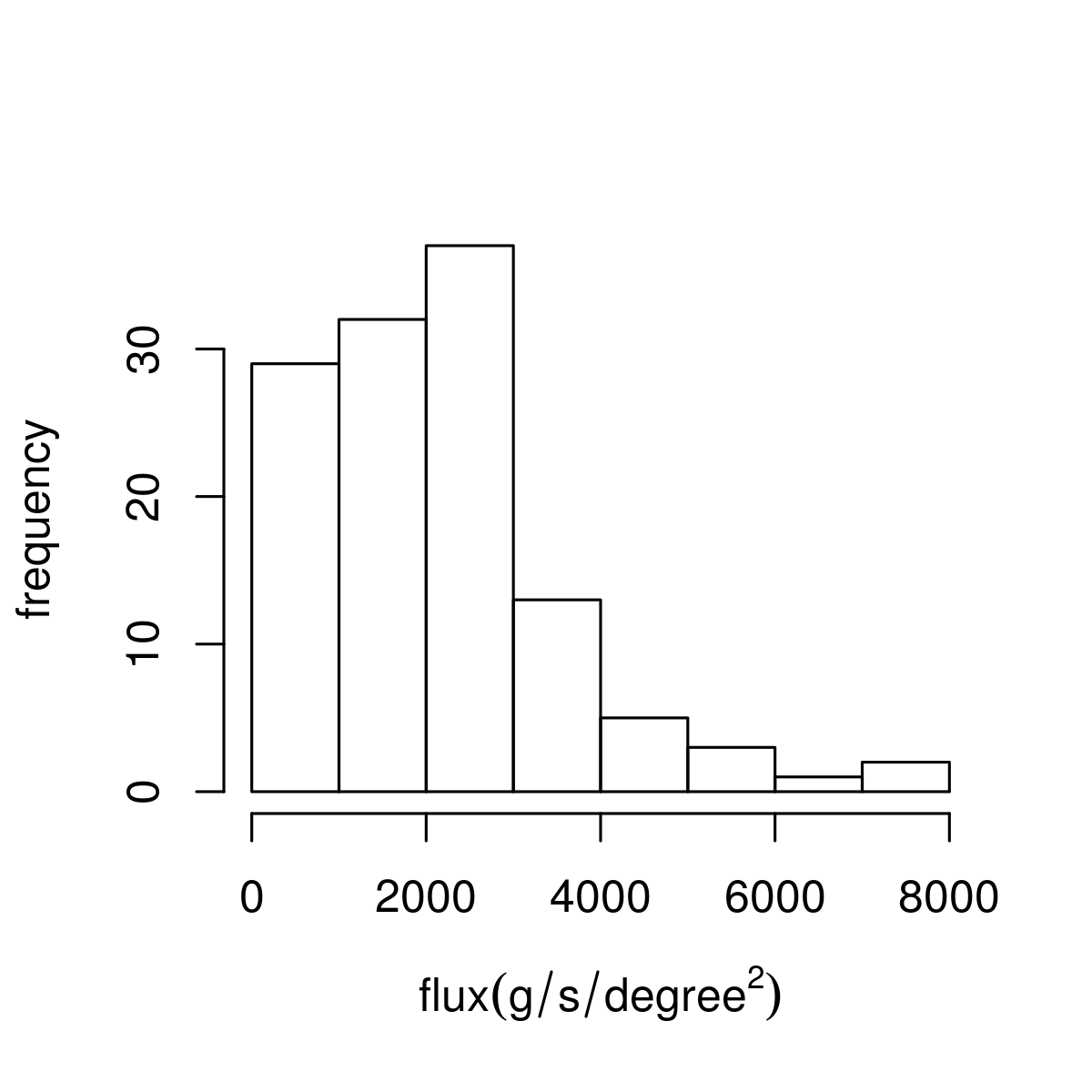}  
\includegraphics[width=3.0in]{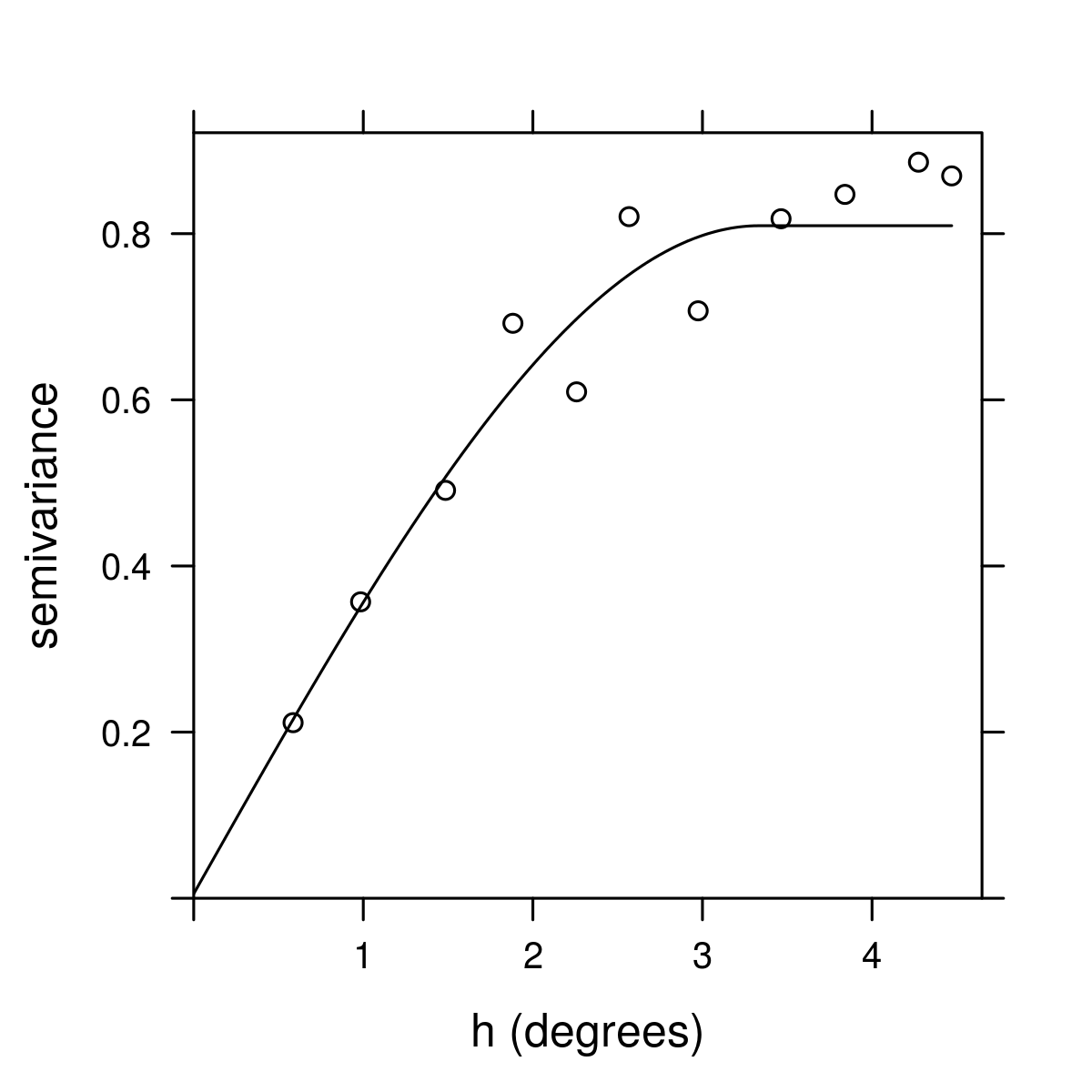}  
	\caption{Left panel: Histogram of total methane flux by grid cell in the UK and Ireland (see Figure \ref{fig:intro}, top-left panel). Right panel: The empirical (open circles) and fitted (solid line) isotropic semivariogram of the methane flux as a function of lag distance in degrees (latitude and longitude). } \label{fig:var_est}
\end{center}
\end{figure}

The flux field is generally spatio-temporal and not isotropic, but here we only seek to capture the dominant spatial dependence using a simple spatial isotropic model; the spatio-temporal discrepancy term $\zeta_t(\svec)$ accounts for any remaining variability. First, we divide the total flux by grid cell in our inventory (shown in Figure \ref{fig:intro}, top-left panel) by the grid cell area in order to obtain flux densities, which we collect into a vector $\Zvec_{inv}$. Then, we let $\Yvec_{inv} \equiv \ln\Zvec_{\inv}$, elementwise. Note that $\Yvec_{inv}$ can be considered dimensionless since, for any constant $k$, $\ln(kZ^i_{inv}) = \ln{k} + \ln Z^i_{inv}$, where $Z^i_{inv}$ is the $i$th element of $\Zmat_{inv}$. A histogram of the elements in $\Zvec_{inv}$ is given in Figure \ref{fig:var_est}, left panel, while the empirical isotropic semivariogram based on $\Yvec_{inv}$ is shown in Figure \ref{fig:var_est}, right panel.

To obtain $\widetilde{C}_{\ff}(\s,\u \mid \varthetab)$, we first fit a semivariogram, $\gamma^o(h\mid \varthetab)$, to the empirical semivariogram shown in Figure \ref{fig:var_est}, where $h = \|\u - \s\|$, and $\varthetab$ is the vector of parameters appearing in the specification of the semivariogram.  Parameter estimates $\hat\varthetab$ were obtained using weighted least squares with \texttt{fit.variogram} in the package \texttt{gstat} with \emph{R Software}  \cite{R,gstat}. We fitted three semivariogram models: the spherical, the exponential, and the Gaussian semivariogram. From these three, the spherical model was the one that gave the best fit in terms of smallest sum-of-squared errors and thus is the one we proceeded with in our analysis. For the spherical semivariogram, $\varthetab \equiv (\sigma^2_1, \sigma^2_2, R)'$, $\sigma^2_1$ is the dimensionless variance of a fine-scale variation component (nugget), $\sigma^2_2$ is the dimensionless variance of a spatially correlated component (partial sill), $R$ is a range parameter in degrees (latitude and longitude), and 
\begin{equation}\label{eq:variogram}
\gamma^o(h\mid \varthetab) \equiv \left\{ \begin{array}{ll}0, & h=0; \\ \sigma^2_1 + \sigma^2_2((3h)/(2R) - (1/2)(h/R)^3),~~~~& 0 < h < R; \\ 
\sigma^2_1 + \sigma^2_2, & h \ge R.  \end{array} \right.
\end{equation}
\noindent The spherical semivariogram has a finite sill, $\sigma_1^2 + \sigma_2^2$, and hence there is a stationary covariance function \cite[][Section 2.3.2]{Cressie_1993},
\begin{equation}
\widetilde{C}_\ff(\svec,\uvec\mid  \varthetab) \equiv \sigma^2_1 + \sigma^2_2 - \gamma^o(\| \uvec - \svec \|\mid \varthetab).\label{eq:cov_fun}
\end{equation}
The parameters corresponding to the fitted semivariogram, shown in Figure \ref{fig:var_est}, right panel, are $\hat\varthetab = (0.0053,0.80,3.3^\circ)'$. The fitted values were used in \eqref{eq:cov_fun} to obtain the covariance function, $\widetilde{C}_\ff(\svec,\uvec\mid  \hat\varthetab)$, which is used to describe the spatial dependence in the log-flux field. We complete the prior specification of the flux field by recalling that $\widetilde{\mu}_f(\cdot) = c$, and setting $c = \sum_{\svec \in D_f^L}Y_\inv(\svec) / |D_f^L| = 7.35$, which is a spatially invariant, dimensionless constant. We verified the adequacy of the fitted lognormal process through a leave-one-out cross-validation study and analysis of the prediction residuals in the log space.

\section{Methods}\label{sec:inference}

Atmospheric trace-gas-inversion problems are computationally demanding and the bivariate model does not fully alleviate this. First, there are four domains to consider, $\mathcal{T}, D^O_m, D^L_m$, and $D^L_f$, the sizes of which can affect the computational efficiency and hence the strategy we employ. Second, even after estimating $\varthetab$ using the inventories, there are still unknown parameters in the discrepancy that need to be estimated; we denote those parameters as $\thetab$. For small problems, a full Markov chain Monte Carlo (MCMC) procedure to sample from the posterior distribution, $[\Yvec_f, \Yvec_m, \thetab \mid  \Zvec_m,  \hat\varthetab]$, may be employed \cite{Ganesan_2015} and, when $|\mathcal{T}|$ grows, exact or approximate algorithms that process the data one-temporal-interval-at-a-time may be used \cite{Carter_1994,Zammit-Mangion_2012}. 

Hierarchical models where prior distributions are placed on the unknown parameters $\thetab$, are known as \emph{Bayesian} hierarchical models (BHMs). As both $|D^L_m|$ and $|D^L_f|$ increase, MCMC schemes may become prohibitive due to the time taken to obtain a single sample from the posterior distribution and due to poor mixing resulting from increased correlations between $\thetab$ and $(\Yvec_f',\Yvec_m')'$. To deal with this, one can implement \emph{empirical} hierarchical models (EHMs, see \cite[][pp.~20--21]{Cressie_2011}) in which the parameters $\thetab$ are treated as fixed unknowns that are estimated using standard maximum-likelihood techniques. MCMC is then used to sample from the empirical predictive distribution, $[\Yvec_f, \Yvec_m \mid  \Zvec_m,  \hat\thetab, \hat\varthetab]$, where the estimate $\hat\thetab$ of $\thetab$ is substituted into the posterior distribution in place of $\thetab$. EHMs do not account for `parameter uncertainty' in the Bayesian sense, and prediction intervals obtained for $[\Yvec_f, \Yvec_m \mid  \Zvec_m,  \hat\thetab, \hat\varthetab]$ tend to be narrower than those obtained using BHMs \cite{Ganesan_2014,Kang_2009}. Nonetheless, EHMs have been shown to give reasonable results at a fraction of the computational cost of BHMs \cite{Sengupta_2013}. We discuss how adjusted intervals may be obtained for EHMs in Section \ref{sec:discussion}.

\subsection{Estimating the remaining parameters}\label{sec:EM}

Since the flux and mole-fraction fields are not directly observed, the likelihood of $\thetab$ is not available in closed form. Maximum-likelihood estimation in this context can be treated as a `missing-data' problem. Missing-data problems appear in all statistical contexts, and they are often solved using iterative methods. One of these methods is the expectation-maximisation (EM) algorithm, given in the celebrated article of \cite{Dempster_1977}. Since its appearance, the EM algorithm has been widely employed for parameter estimation in empirical hierarchical spatio-temporal models (e.g., \cite{Dewar_2009}), where the latent process is treated as missing data.



The EM algorithm for estimating $\thetab$ is relatively straightforward to implement. Define the complete-data likelihood, $L_c(\thetab) \equiv [\Yvec_f,\Yvec_m,\Zvec_m \mid  \thetab, \hat\varthetab]$. Thus, $L_c(\thetab)$ inherits its randomness from $\Yvec_f,\Yvec_m$, and $\Zvec_m$. The EM algorithm is defined iteratively through: (i) the E-step, in which the function
\begin{equation*}
Q(\thetab\mid  \thetab^{(i)}) \equiv \E(\ln L_c(\thetab) \mid \Zvec_m , \thetab^{(i)}, \hat\varthetab),
\end{equation*}
is found for some `current' estimate $\thetab^{(i)}$, and (ii) the M-step, which computes
\begin{equation}\label{eq:Mstep}
\thetab^{(i+1)} = \argmax_{\thetab} Q(\thetab\mid  \thetab^{(i)}).
\end{equation}
The algorithm is iterated until convergence (convergence is assessed by monitoring the change in the elements of $\thetab^{(i)}$ for successive values of $i$). For a comprehensive treatment of the EM algorithm, see \cite{Mclachlan_1997}.

A slight modification to the standard EM algorithm is necessary in this setting: Because $\Yvec_f$ is lognormal, the E-step cannot be carried out analytically. However, since $\zeta_t(\svec)$ is a Gaussian process, it turns out that we only need to compute the conditional expectation and the conditional covariance of $(\Yvec_f,\Yvec_m) \mid  \Zvec_m , \thetab^{(i)},\hat\varthetab$ (see \eqref{eq:Psi} in \ref{sec:Psisec}). We obtain these by finding the mode and the curvature of the log-density (thus invoking a Laplace approximation; see \cite[][p.~213]{Bishop_2006}). Formulas required for the E-step are given in \ref{sec:Estep}. The resulting (approximate) algorithm is referred to as a Laplace-EM algorithm in \cite{Sengupta_2013}.

 For the models we consider, the optimisation \eqref{eq:Mstep} can be carried out using gradient descent. Since computing the derivatives of $Q(\thetab\mid \thetab^{(i)})$ can be quite costly, we suggest halting the gradient descent prematurely for the first few M-steps in order to decrease computational cost. The resulting `generalised EM algorithm' \citep[][p.~84]{Mclachlan_1997} converges to a stationary point $\hat\thetab$ (when the E-step is exact), although it requires more iterations until convergence.

\subsection{Computationally efficient models for the discrepancy term}\label{sec:models}

As discussed in Section \ref{sec:bivariate}, one reason to employ a bivariate model is the flexibility it gives in choosing a model for the discrepancy term. Clearly, one should choose a model for the discrepancy that allows it to scale well with both $|\mathcal{T}|$ and $|D^L_m|$. In this article we employ a separable space-time structure (e.g., \cite{Ganesan_2015,Gneiting_2006}) that is computationally efficient and is a reasonable choice in the face of lack of structural information on the discrepancy term $\zeta_t(\svec)$; see Section \ref{sec:discussion} for further discussion of this choice.

A  separable covariance function is one that can be characterised through
\begin{equation*}
\cov(\zeta_t(\svec),\zeta_{t'}(\uvec) \mid \thetab) \equiv \sigma^2_\zeta\rho_s(\svec,\uvec\mid \thetab)\rho_t(t,t'\mid \thetab),
\end{equation*}
where $\rho_s(\cdot,\cdot \mid \thetab)$ is a spatial correlation function and $\rho_t(\cdot,\cdot \mid \thetab)$ is a temporal correlation function. In this work, we restrict our attention to the following correlation functions:
\begin{align*}
\rho_s(\svec,\uvec\mid  d) &\equiv \exp(-\| \uvec - \svec \| / d);\quad d>0, \\
\rho_t(t,t'\mid  a) &\equiv a^{|t' - t|}; \quad |a| < 1,
\end{align*}
where $d$ and $a$ are range-like parameters that determine the correlation length of the spatial and temporal components, respectively (we refer to $d$ as a correlation e-folding length).  The derivatives of these correlation functions with respect to $d$ and $a$ can be easily computed for the M-step. Recall that, in our model, the unknown parameters $\thetab$ are all associated with the discrepancy term's covariance function; therefore, $\thetab \equiv (\sigma^2_\zeta,a,d)'$.


Separability implies that 
\begin{align*}
\Sigmamat_\zeta(\thetab) &\equiv (\cov(\zeta_t(\svec),\zeta_{t'}(\uvec) \mid \thetab) : \svec,\uvec \in D^L_m;\, t,t' \in \mathcal{T}) \\
&= \sigma^2_\zeta\Rmat_{\zeta,t}(a) \otimes \Rmat_{\zeta,s}(d),
\end{align*}
where $\otimes$ is the Kronecker product, $\Rmat_{\zeta,t}(a)$ is a correlation matrix of size $|\mathcal{T}| \times |\mathcal{T}|$ and $\Rmat_{\zeta,s}(d)$ is a correlation matrix of size $|D_m^L| \times |D_m^L|$. One can take advantage of the Kronecker product to greatly reduce the memory and computational requirements of the E-steps and M-steps.  Formulas required for the M-step for this model are given in \ref{sec:Mstep}.

\subsection{Taking advantage of the closed-form gradient: Hamiltonian Monte Carlo}\label{sec:mcmc}

Recall that with an EHM, we first estimate the parameters using, for example, the EM algorithm. Then, after fixing them at the EM-estimate, we carry out inference on the bivariate fields using MCMC. This section describes the second, MCMC, stage for sampling from the empirical predictive distribution, $[\Yvec_f,\Yvec_m \mid  \Zvec_m, \hat\thetab, \hat\varthetab]$.

The most popular MCMC algorithm in use is a random-walk Metropolis algorithm, where one uses a symmetric density to propose new states in the Markov chain. Metropolis algorithms do not take into account the `shape' of the density and, as a result, the optimal acceptance rate typically results in chains that are highly correlated. We can improve on the Metropolis sampler by using the gradient of the posterior density, which in our case can be computed analytically. 
The gradient is used to reduce the probability of proposing samples in the direction where the probability density drops off steeply, and to increase the probability in regions where the probability density is relatively smooth. Hamiltonian (or hybrid) Monte Carlo (HMC) is an MCMC method that takes into account gradient information and can thus be used to propose states that might be distant, in probability space, from the current sample and yet have a small chance of rejection. In practice, a trajectory for the samples is simulated using Hamiltonian dynamics \cite{Duane_1987}. 

A Hamiltonian system is described by a position vector and a momentum vector. In HMC, the position vector corresponds to the current sample, while the momentum vector is randomly sampled from a pre-specified distribution. Hamilton's equations are then used to simulate the trajectory of the current sample, under the random momentum vector, for a certain time interval. The potential-energy surface used is the negative log-density, and thus the trajectory of the sample can be thought of as a particle rolling inside a surface characterised by the probability density of interest. In practice, Hamilton's equations are discretised in time using a suitable scheme like the leapfrog method \cite{Neal_2011}, with a given discretisation interval $\Delta$. Then the number of time steps is $L = $ (time interval)/$\Delta$. Choice of $\Delta$ and $L$ are critical for good performance of the MCMC scheme.

In general, $\Delta$ needs to be set such that the narrow, possibly curving valleys in the negative log-density can be explored, while $L$ needs to be set such that the time interval, $L\Delta$, is large enough for the proposal to be treated as independent from the current sample. This is no simple task, however by transforming the variables to have roughly unit scale, we can explore values of $L$ and $\Delta$ for a suitable acceptance rate ($\approx 60\%$) such that $L\Delta \simeq 1$ \cite{Neal_2011}. The quantity $\Delta$ can also be randomised in order to minimise the chance of obtaining periodicity in the trajectories; see \cite{Neal_2011} for further implementation details. 

Since $\Yvec_m$ is conditionally Gaussian, conditional on $\Yvec_f, \hat\thetab,$ and $\hat\varthetab$, we can marginalise out $\Yvec_m$ from $[\Yvec_f,\Yvec_m \mid  \Zvec_m, \hat\thetab, \hat\varthetab]$ and use HMC to sample $\{\Yvec_f^{(i)}\}$ directly from $[\Yvec_f \mid  \Zvec_m, \hat\thetab,\hat\varthetab]$.  For each $\Yvec_f^{(i)}$, we then sample $\Yvec_m^{(i)}$ from $[\Yvec_m \mid   \Yvec_f^{(i)}, \Zvec_m, \hat\thetab,\hat\varthetab]$, which is Gaussian. This sampling scheme is known as a collapsed Gibbs sampler \cite{Dyk_2008}.  The log-density and the gradient for the collapsed distribution (i.e., $[\Yvec_f \mid  \Zvec_m, \hat\thetab,\hat\varthetab]$) can be derived in a very similar way to the Laplace-approximated E-step (\ref{sec:Estep}), and thus they are not detailed here. 

\section{Results}\label{sec:results}

In this section, we implement the model of Section \ref{sec:model} and the inference techniques discussed in Section \ref{sec:inference} in two settings. First, in Section \ref{sec:sim} we use simulated datasets on a one-dimensional `toy' problem. Second, in Section \ref{sec:uk} we apply the method to inferring methane emissions from mole-fraction observations in the UK and Ireland, as described in Section \ref{sec:Intro}.

\subsection{Study with simulated data}\label{sec:sim}

In the simulation study we simulate the flux from some known prior distribution, and the interaction function, which describes the SRR, is synthesised from a spatio-temporal model. For consistency with the analysis of methane emissions given in Section \ref{sec:uk}, we assume that the mole fractions obtained using the simulated flux field and the SRR are in ppb, that the spatial-length scales are in degrees, and that each time step corresponds to a temporal interval of 2 h. In order to facilitate the analysis and visualisation of the results, we carry out these simulations in one-dimensional space. Specifically, the domain of interest is $D \equiv [-10^\circ,10^\circ]$, and $D^L_f \equiv \{-9.9^\circ, -9.7^\circ,\dots,9.9^\circ \}$, so that $|D^L_f| = 100$. The lognormal spatial process that defines the flux field has $\widetilde\mu_f(s) \equiv 5$ and $\widetilde{C}_{\ff}(s,u\mid  \varthetab)$ given by a spherical, isotropic covariance function, with parameters $\varthetab$ set equal to those estimated in Section \ref{sec:data-feature}.

\begin{figure}[!t]
\begin{center}

\includegraphics[width=6in]{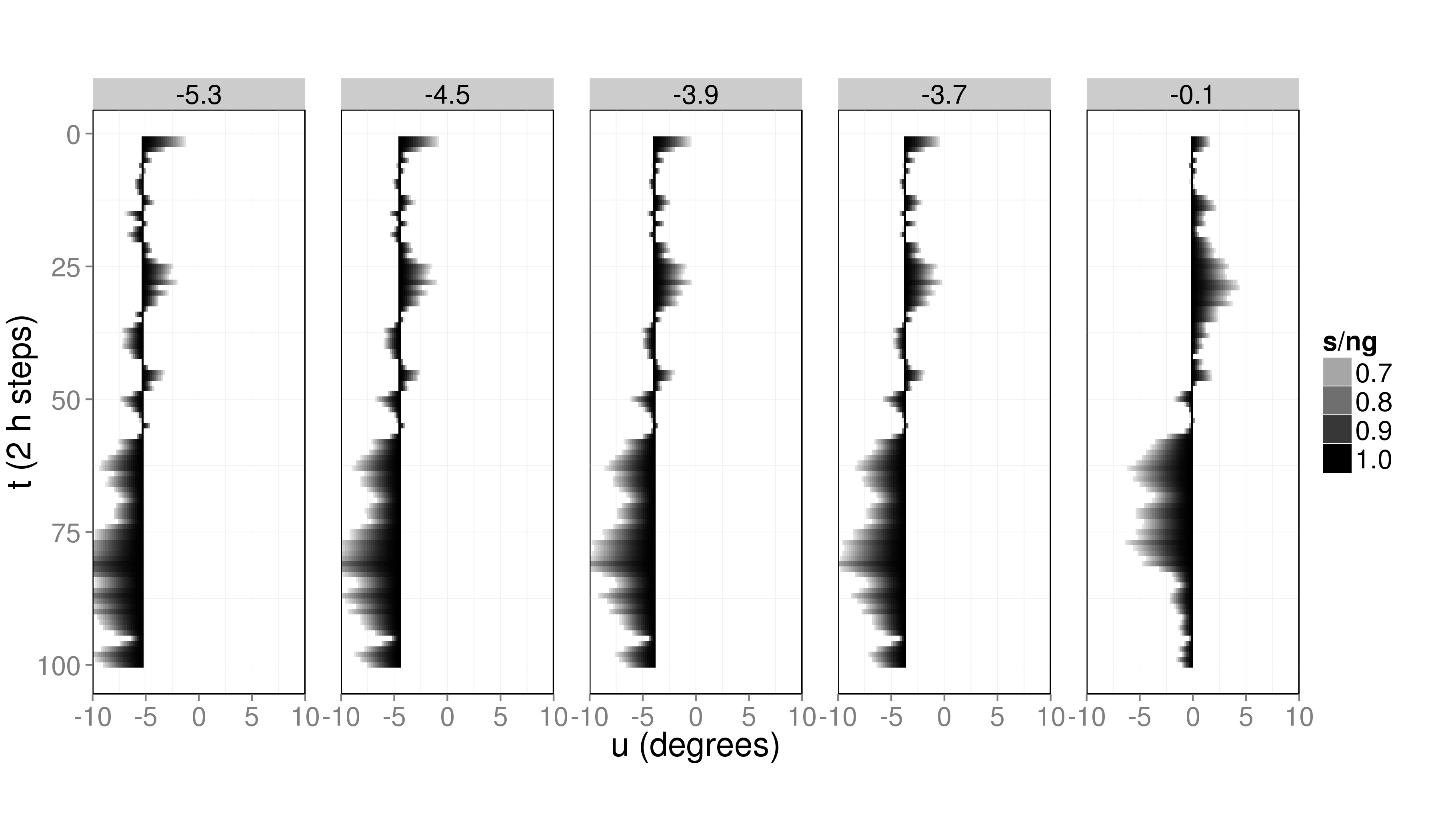}  

\vspace{0.2in}

\includegraphics[width=6in]{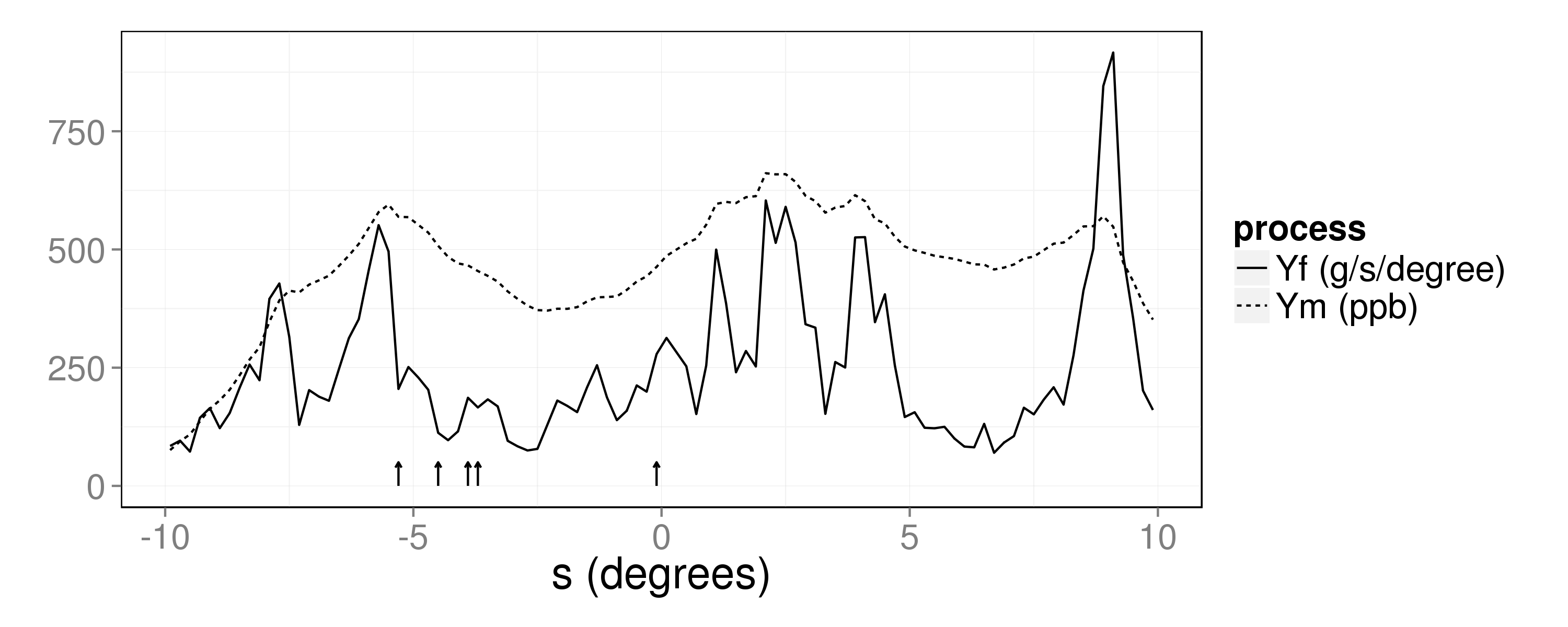} 
	\caption{Top panel: The source-receptor relationship $b_t(s,u)$ synthesised at five observation locations $s \in D^O_m = \{-5.3^\circ, -4.5^\circ,-3.9^\circ,-3.7^\circ,-0.1^\circ\}$. Note that $b_t(s,u) = 0$ for $u > 4.3^\circ$. Bottom panel: A sample realisation of the flux field (solid line), the resulting time-averaged mole-fraction field (dashed line), and the five observation locations (arrows).} \label{fig:B}
\end{center}
\end{figure}

We synthesised an SRR that reflects something that would be obtained from a numerical model such as NAME. That is, we use an SRR characterised by a plume originating at $u = s$ with a particular orientation that decays slowly with $|u-s|$ (see Figure \ref{fig:intro}, top-right panel) and varies smoothly in time. We achieved this by simulating a flow process through a truncated normalised Gaussian density with spatio-temporally varying scale. Specifically, we first simulated a parameter $\upsilon_t(s)$ from a Gaussian process with separable spatio-temporal covariance structure, and then we defined
\begin{equation}\label{eq:bt}
b_t(s,u\mid  \upsilon_t(s)) \equiv \exp\left(-\frac{(u-s)^2}{2\upsilon_t(s)^2}\right)I\big(|u-s| < |\upsilon_t(s)|\big)J(s,u),
\end{equation} 
\noindent where 
\begin{equation*}
J(s,u) \equiv \left\{ \begin{array}{ll} I[(u-s) \ge 0]; & \upsilon_t(s) \ge 0, \\
                                        I[(u-s) \le 0]; & \upsilon_t(s) < 0,
\end{array}\right.
\end{equation*}
and $I(\cdot)$ is the indicator function. In \eqref{eq:bt}, the exponential function describes a bell-shaped curve centred at $u = s$, where the indicator function truncates this curve at $u = s \pm \upsilon_t(s)$. The third term, $J(s,u)$, then truncates the bottom half of the function if $\upsilon_t(s) \ge 0$ and the upper half otherwise. A plot showing $\{b_t(s,u): u \in D^L_f;\, t \in \mathcal{T}\}$ at five locations $s$ (that coincide with the observation locations, $D_m^O$,  to be described later), where $\mathcal{T}\equiv\{1,2,\dots,T\}, T = 100$ steps, is shown in Figure \ref{fig:B}, top panel. 

The SRR was subsequently used to construct the matrix $\Bmat_{B,t}$ defined by \eqref{eq:BMt}, with $A(\uvec) = 0.2^\circ$. For the purposes of simulation, we set the true-discrepancy parameters to $\sigma_\zeta = 50$ ppb, $a = 0.8$, and $d = 1^\circ$, so that $\thetab = (2500$ ppb$^2, 0.8, 1^\circ)'$.  We also set $\E(\zeta_t(s)) = 0$, thereby assuming that any background mole fraction has been correctly removed \emph{a priori}. Using these parameters and assumptions, we then simulated the flux field from \eqref{eq:ln1} and \eqref{eq:ln2} and the mole-fraction field from \eqref{eq:mv}. This simulation was done only once, to give us a simulated dataset where the true fields are known. In Figure \ref{fig:B}, bottom panel, we show the true lognormal flux field and the true time-averaged mole-fraction field with $D^L_m \equiv D^L_f$.

We first considered the typical scenario of when we have only a few measurement stations where mole-fraction data are collected. For example, in Section \ref{sec:uk}, we have only four stations in the UK and Ireland. In this simulation study in one-dimensional space we considered five stations, shown by the arrows in the bottom panel of Figure \ref{fig:B},  whose locations $D^O_m$ were obtained by sampling randomly from $D^L_f$. Data were generated at these locations using \eqref{eq:obs}, and we assumed that data for all time steps were available (i.e., $D^O_{m,t} = D^O_m;\, t \in \mathcal{T}$). Recall that $T=100$ time steps, which corresponds to about 8 days of 2-hourly observations. We assume that we are interested in predicting the mole-fraction field at the observation locations and, for illustration, at another location $s = 0.3^\circ$, and hence we set $D^L_m = D^O_m \cup \{0.3^\circ\}$.


\newcommand*{\medcup}{\mathbin{\scalebox{1.5}{\ensuremath{\cup}}}}%
The function $b_t(s,u)$ is a measure of how sensitive $Y_{m,t}(s)$ is to flux at $u$. It can be problematic, in ill-posed problems such as this one, to obtain inferences for locations $u$ when the sensitivity at all measurement sites $D^O_{m,t}$ is zero (or approximately zero) for $t \in \mathcal{T}$, since the influence of the flux field at these locations on the observed mole fractions is zero (or negligible). Without  assuming a spatial process for the flux field, these regions would not even be identifiable, and hence they can be justifiably removed from the model. We therefore re-defined the flux domain $D^L_f$ to exclude these regions. Formally, we can define $D_t(s) \equiv \{u : b_t(s,u) > 0\}$ and re-define $D^L_f \equiv \{-9.9^\circ,-9.7^\circ,\dots,9.9^\circ\} \cap \left(\medcup_{ s \in D^O_m, t \in \mathcal{T}}D_t(s)\right).$


In order to run the Laplace-EM algorithm, we chose the starting value, $\thetab^{(0)} \equiv (1000$ ppb$^2,0.2,0.2^\circ)'$, and commenced the gradient descent at $\Yvec_f = \exp(\hat\mu_f + 0.5\hat\sigma^2_f)\oneb$ (where $\hat\mu_f = 5$ and $\hat\sigma^2_f \equiv \widetilde{C}_{\ff}(s,s \mid \hat\varthetab))$ and $\Yvec_m = \Zvec_m$. Using 50 gradient descents at each M-step, the Laplace-EM algorithm converged in 40 iterations to $\hat\thetab = (2000$ ppb$^2,0.71, 0.77^\circ)'$ (convergence was assessed visually).  Recall that the true value of the parameters were $\thetab = (2500$ ppb$^2, 0.8, 1^\circ)'$. We can expect the accuracy of these estimates to increase with the length of the observation record, $T$. Hence, when using this method in temporal blocks (e.g., \cite{Ganesan_2015,Ganesan_2014}), it is recommended to use large temporal intervals; for example in the case study in Section \ref{sec:uk} we employ a three-month block with $T \approx 1000$ time steps. 

\begin{figure}[!t]
\begin{center}
\includegraphics[width=6.0in]{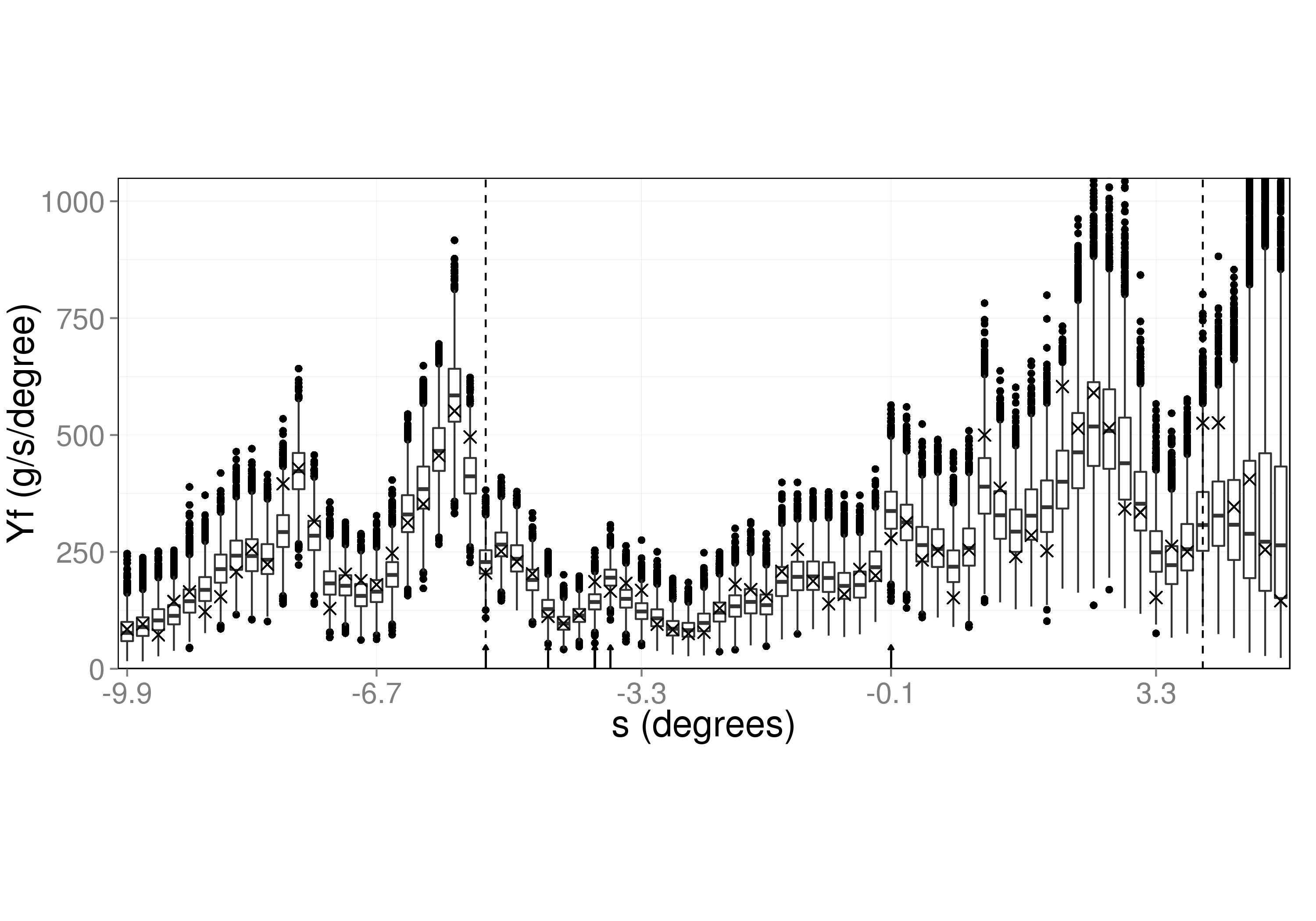}  
	\caption{Samples from the posterior distribution of the lognormal flux field obtained using Hamiltonian Monte Carlo (HMC). The boxes denote the interquartile range, the whiskers extend to the last values that are within 1.5 times the interquartile range from the quartiles, and the dots show the samples that lie beyond the end of the whiskers. The crosses denote the true (simulated) fluxes, and the arrows along the horizontal axis denote the locations of the measurements. The vertical dashed lines show the spatial locations analysed in Figure \ref{fig:Lap_vs_mcmc}. Since the observed mole fractions are insensitive to flux at $\{s : s > 4.3^\circ\}$, these locations were excluded from the model.} \label{fig:Slice}
\end{center}
\end{figure}

We substituted the EM-estimated $\hat\thetab$ for $\thetab$ into $[\Yvec_f,\Yvec_m \mid \Zvec_m,\thetab,\hat\varthetab]$ and sampled 10,000 times from the (empirical) posterior distribution of the flux field using the HMC sampler of Section \ref{sec:mcmc} with $L = 10$ steps and $\Delta \in [0.066,0.068]$; this resulted in an acceptance rate of 57\%. Box plots showing inferences for the flux field from the HMC sampler (after discarding the first 1,000 samples for burn-in), together with the true flux field (crosses) and the observation locations (arrows on the horizontal axis) are shown in Figure \ref{fig:Slice}. Notice how the uncertainty increases as distance away from the nearest observation location increases, as expected, but also notice how the true fluxes conform with our posterior inferences. 

\begin{figure}[!ht]
\begin{center}
\includegraphics[width=3in]{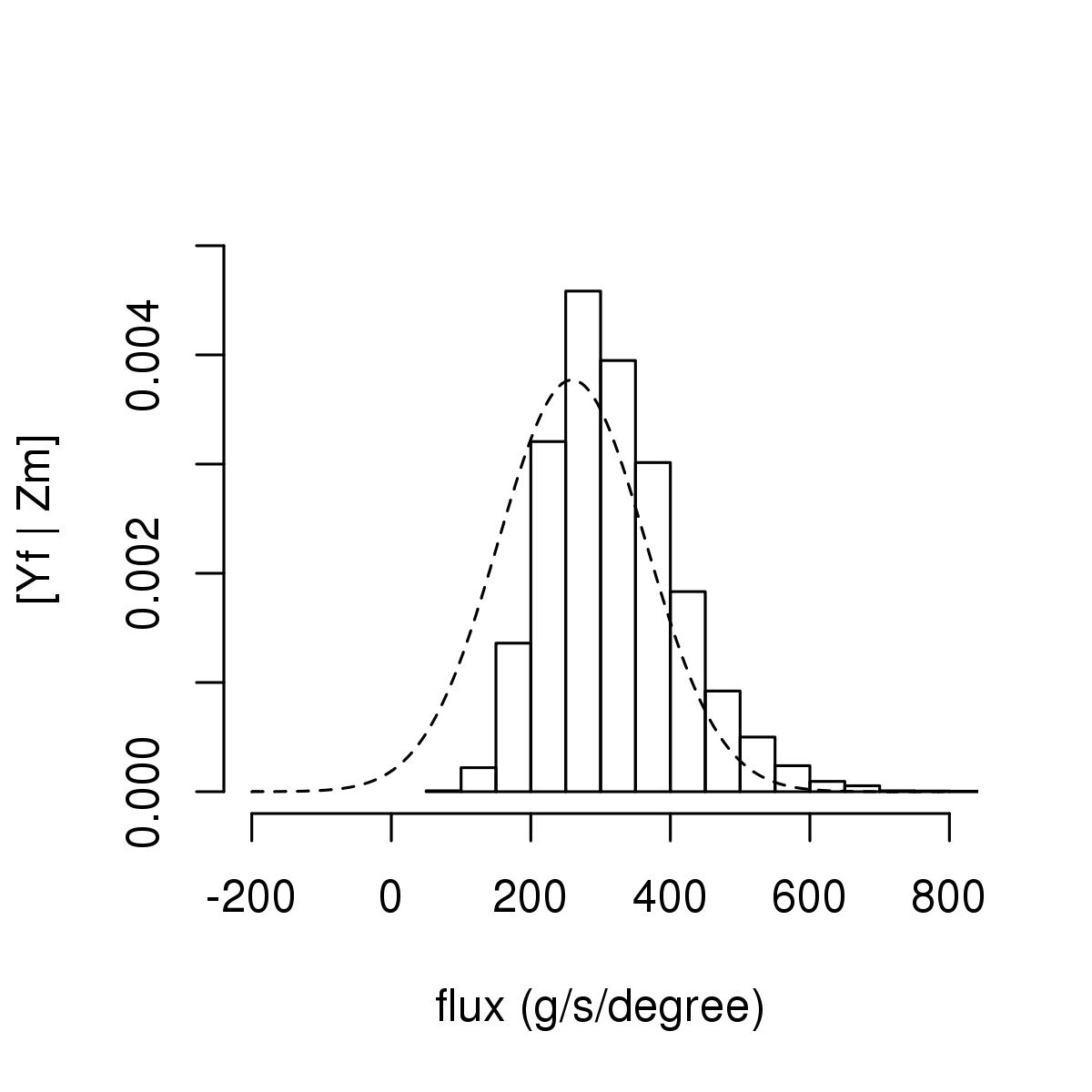}  
\includegraphics[width=3in]{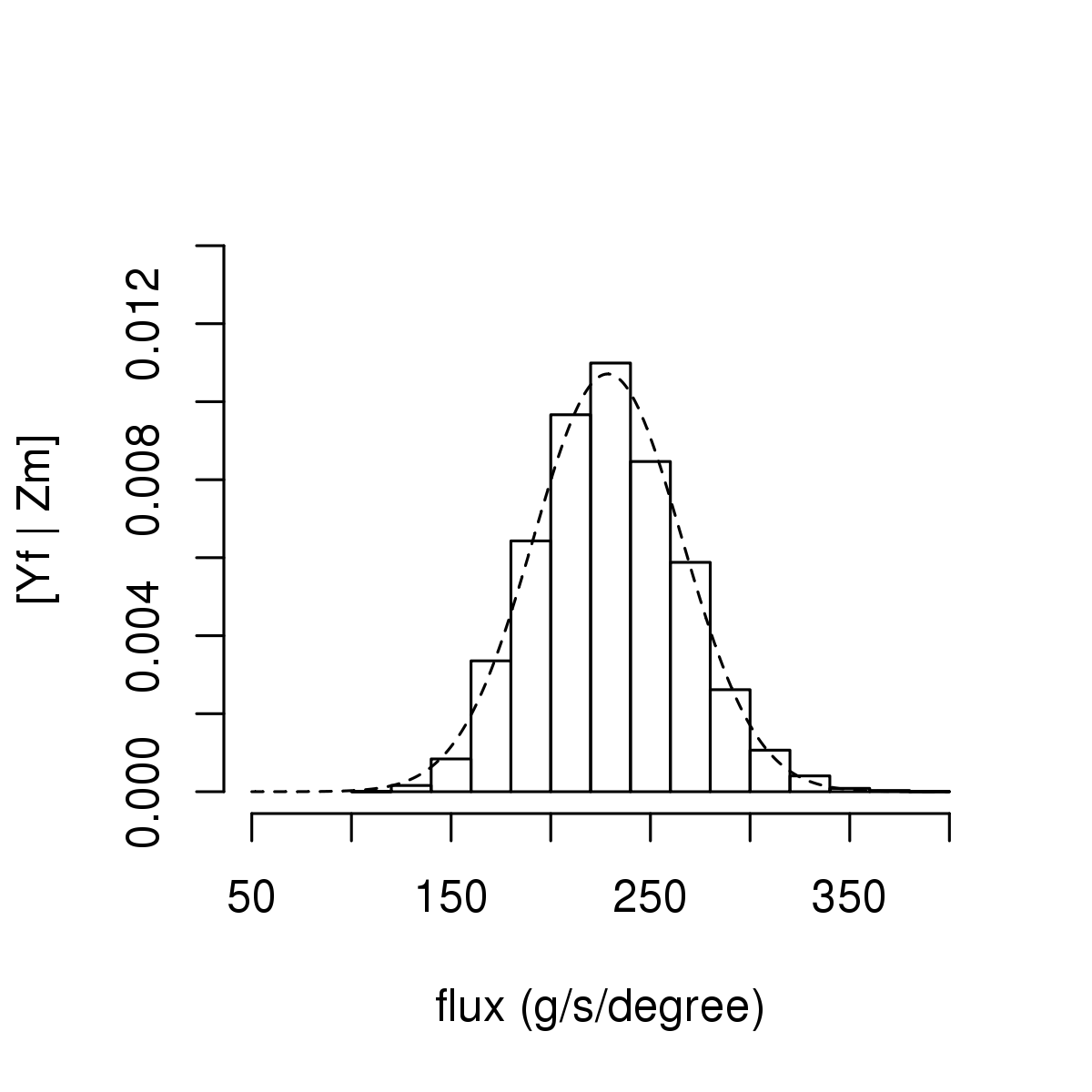}
	\caption{Laplace approximation (dashed line) and a histogram of the empirical posterior distribution from the MCMC samples for the flux at $s = 3.9^\circ$ (left panel) and $s = -5.3^\circ$ (right panel).} \label{fig:Lap_vs_mcmc}
\end{center}
\end{figure}

The Laplace approximation is used for parameter estimation and not for prediction of the fields, although predictions could be provided as by-products of the E-step at convergence of the EM algortihm. The reason we do not use the predictions from the Laplace approximation is that, while correctly locating the mode in the posterior distribution, these predictions do not account for skewness, something that is expected from a lognormal process. The difference between the prediction from the Laplace approximation and that from the HMC sampler at $s = 3.9^\circ$ (far from an observation location) is shown in Figure \ref{fig:Lap_vs_mcmc}, left panel. We find that this difference is less marked close to an observation location, since the posterior distribution is less sensitive to the prior distribution in these regions (e.g., at $s = -5.3^\circ$); see Figure \ref{fig:Lap_vs_mcmc}, right panel.


One of the contributions of this article is to introduce a lognormal-spatial model for the flux field, for trace gases where fluxes are positive. We show the importance of capturing the spatial characteristics of the flux field by comparing our results obtained with spatial range $R = 3.3^\circ$, to results we obtained with a misspecified, non-spatial model where $R = 0.0^\circ$ in \eqref{eq:variogram}. To assess the `goodness-of-prediction' for both models we use the root-average-squared prediction error:
\begin{equation}\label{eq:S1f}
S_{1,f} = \left[ \frac{1}{|D^L_f|} \sum_{i = 1}^{|D^L_f|}\left(Y_{f,i} - \hat{Y}_{f,i} \right)^2  \right]^{1/2},
\end{equation}

\noindent where $Y_{f,i}$ is the $i$th element of the vector $\Yvec_f$ and $\hat Y_{f,i}$ is the posterior expectation of $Y_{f,i}$ assuming a spatial range that may be different from the true range. The lower the value for $S_{1,f}$, the better the predictive performance in terms of pointwise prediction. To assess the distributional accuracy of the forecast, we use 
\begin{equation}\label{eq:S2f}
S_{2,f} = \left[ \frac{\frac{1}{|D^L_f|} \sum_{i = 1}^{|D^L_f|}\left(Y_{f,i} - \hat{Y}_{f,i} \right)^2}{\frac{1}{|D^L_f|} \sum_{i = 1}^{|D^L_f|} \sigma^2_{Y_{f,i}}}  \right]^{1/2},
\end{equation}
\noindent where $ \sigma^2_{Y_{f,i}}$ is the posterior variance of $Y_{f,i}$. Clearly, $S_{2,f}$ should be close to 1 if uncertainty is correctly captured.

\begin{center}
\begin{table}\caption{Statistics $S_{1,f}$, $S_{2,f}$, and $S_{1,m}^{0.3}$ for the simulated case studies using a lognormal uncorrelated spatial process ($R=0.0^\circ$) and a lognormal correlated spatial process ($R = 3.3^\circ$)}
\vspace{0.1in}
\centering \begin{tabular}{cc} 
$R$ (degrees)\\ \hline
$0.0$\\
$3.3$
\end{tabular} \begin{tabular}{cx{1cm}cx{3cm}cx{10cm}}
$S_{1,f}$ (g s$^{-1}$ degree$^{-1}$) & $S_{2,f}$ & \qquad$S_{1,m}^{0.3}$ (ppb)\\ \hline
97       & 0.74 & \qquad 32    \\
62       & 0.83 & \qquad 32
\end{tabular} \label{tab:stats}
\end{table}
\end{center}

In Table \ref{tab:stats}, we show the statistics \eqref{eq:S1f} and \eqref{eq:S2f} obtained under uncorrelated (i.e., misspecified) and correlated (i.e., correctly specified) flux-field assumptions. The statistic $S_{1,f}$  for the model with the spatially correlated flux was 62 g s$^{-1}$ degree$^{-1}$, while for the uncorrelated model (with $R = 0.0^\circ$), $S_{1,f}$ was 97 g s$^{-1}$ degree$^{-1}$. Assuming an uncorrelated flux field results in a considerable reduction in performance, caused mainly by poor observability of the flux field at $s > 0.0^\circ$ (plot not shown). Because of such regions, it is important to capture the spatial correlations of the unobserved fields. Also, the statistic $S_{2,f}$ is closer to 1  when $R = 3.3^\circ$ than when $R = 0.0^\circ$. A value lower than 1 is indicative of under-confidence (an overall posterior variance which is too large); in our case, $R = 0.0^\circ$ represents a model misspecification, and we see the consequence in the low value of $S_{2,f}$ (namely, 0.74).

\begin{figure}[!t]
\begin{center}
\includegraphics[width=6.0in]{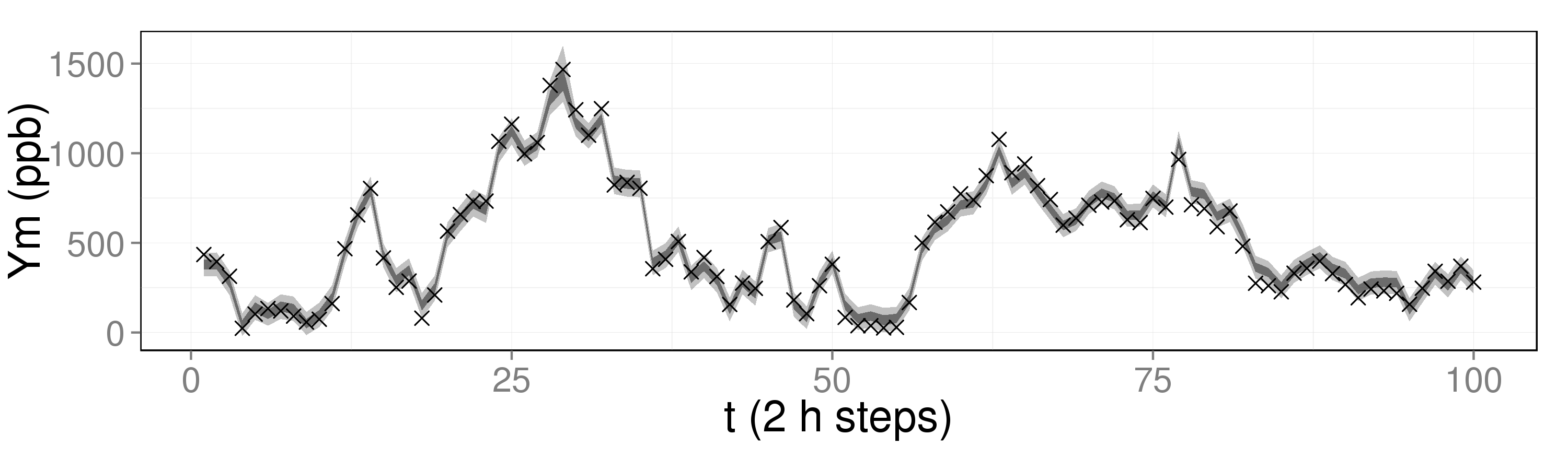}  
	\caption{Distributions of mole fraction at $s = 0.3^\circ$ and $t \in \mathcal{T}$, following Gibbs sampling. The dark and light shadings denote the interquartile and the 5th--95th percentile ranges, respectively. The crosses denote the true (simulated) mole fractions.} \label{fig:MF}
\end{center}
\end{figure}

Another advantage of the proposed approach is the ease with which we can infer the mole-fraction field anywhere in the domain. In Figure \ref{fig:MF}, we show the distributions of the mole fractions at every time point at $s = 0.3^\circ$, which are consistent with the true mole fraction values. A statistic similar to \eqref{eq:S1f} for evaluating the mole-fraction `goodness-of-prediction' at $s = 0.3^\circ$ is
\begin{equation}\label{eq:S1m}
S_{1,m}^{0.3} = \left[ \frac{1}{|\mathcal{T}|} \sum_{t = 1}^{|\mathcal{T}|}\left(Y_{m,t}^{0.3} - \hat{Y}_{m,t}^{0.3} \right)^2  \right]^{1/2},
\end{equation}
\noindent where $Y_{m,t}^{0.3}$ is the element of the vector $\Yvec_{m,t}$ corresponding to the location $s = 0.3^\circ$, and $\hat Y_{m,t}^{0.3}$ is the posterior expectation of $Y_{m,t}^{0.3}$ assuming different values of $R$. Interestingly, unlike with $S_{1,f}$, the statistic \eqref{eq:S1m} was found to be insensitive to $R$; see Table \ref{tab:stats}. This reflects a well-known consequence of the ill-posed nature of the problem: There are several `plausible' flux fields that can reproduce the observed mole fractions.

Next, we show that the bivariate approach scales well with dataset size by considering the case where $|D_m^O| = 1000$. The locations of the 1000 observations were obtained by uniform sampling within the domain, and the observations were generated using \eqref{eq:obs}. Since the number of time steps is $T=100$, we have 100,000 observations in all, far more than can be handled efficiently with the standard univariate approach based on \eqref{eq:uv}. This problem becomes tractable by choosing $D^L_m$ such that $|D^L_m| \ll |D_m^O|$, which is possible only when using a bivariate model. Here we set $D^L_m = D^L_f$ and, since $|D^L_f| = 100$, we have 10,000 space-time locations at which to infer the mole-fraction field, which is an order of magnitude less than $|D_m^O|$. 
The Laplace-EM algorithm converged in 5 iterations to $\hat\thetab = (2300$ ppb$^2$, 0.81, 0.89$^\circ)'$. As expected, these estimates are more accurate due to the increased number of observations (recall that the true parameters were $\thetab = (2500$ ppb$^2, 0.80, 1.00^\circ)'$). Increased accuracy was also noted in the posterior mode of the flux field, which was virtually identical to the true flux field at all locations in $D^L_f$.  

Although it seems relatively straightforward to adopt this separable geostatistical model with large datasets, since both $\Rmat_{\zeta,t}$ and $\Rmat_{\zeta,s}$ are dense, one cannot use it when either $|\mathcal{T}|$ or $|D^L_m|$ become larger than, say, 2000. One way to remedy this limitation is to employ sparse inverse covariance matrices or reduced-rank covariance matrices instead; we provide further discussion on one such approach in Section \ref{sec:discussion}.

Reproducible code for this simulation study is available from \url{https://github.com/andrewzm/atminv}.

\subsection{Inference on methane emissions in the UK and Ireland}\label{sec:uk}

In this section we present a case study that analyses real data on mole fraction in the UK and Ireland from the four stations shown in Figure \ref{fig:intro}, top-right panel, and we infer the discretised flux and mole-fraction fields. 

\subsubsection{Data, preprocessing, and the source-receptor relationships}\label{sec:preprocess}

Methane observations, available in ppb, were made at four sites: Mace Head, Ireland (MHD), Ridge Hill, England (RGL), Tacolneston, England (TAC) and Angus, Scotland (TTA). These sites are from the UK Department of Energy and Climate Change (DECC) observation network \cite{DECC}. In this analysis, the period January--March 2014 was used. The SRR was evaluated at the observation locations using the UK Met Office's Numerical Atmospheric-dispersion Modelling Environment (NAME) \cite{Jones_2007}, which is a Lagrangian Particle Dispersion Model (LPDM) that was run backwards for 30 days. The spatial domain extended from $-14^\circ$ to 31$^\circ$ E and 36$^\circ$ to 66$^\circ$ N at a grid resolution of 0.352$^\circ$ lon by 0.234$^\circ$ lat, thus covering the UK and a large part of continental Europe with a 128 $\times$ 128 grid. For a complete description of the measurement procedure,  ancillary information about the measurement sites, and a description of NAME and its application to prediction of UK and Ireland emissions, see \cite{Ganesan_2015}.


Since the LPDM only simulates the effect of emissions from the previous 30 days before measurement, a substantial `background' level of methane exists in the observed mole fractions that would not be accounted for by the LPDM. This background depends on the direction of the origin of the air mass due to, for example, the N-S latitudinal gradient or the vertical gradient. We used the following simple procedure to cater for the background. First, using NAME, we identified which observations were influenced by southerly transport or were influenced by the upper atmosphere due to vertical transport by tracking  the directions that particles exited the NAME domain. Due to the complexities involved in estimating these backgrounds, these observations were removed from the dataset. Second, winds that were westerly in origin and passed over the Atlantic ocean were considered to be relatively free of regional emissions. Hence, this background, due to air masses originating in the west, can be clearly observed at Mace Head in Ireland. We used the 5th percentile of the extant mole fraction observations at Mace Head as a crude estimate of the background for all observations. Third, we subtracted this background estimate from the extant observations at all sites. In addition to removing data for background purposes, observations that could have significant influence from sub-grid scale processes (defined as `local' processes), also identified using NAME, were removed. These data corresponded to times when there was a significant sensitivity to the nine grid cells surrounding the measurement point (i.e., when air was more likely to be stagnant).


To restrict our attention to land areas in the UK and Ireland, we used the methane flux inventory, together with NAME, to subtract contributions from the rest of Europe, including the sea territories of the UK and Ireland. Thus, we have final, extant observations of methane mole fraction, corrected for background and due to methane emissions only in the UK and Ireland land areas. The flux inventory and the NAME outputs were aggregated to a coarser resolution by grouping 2 $\times$ 2 grid cells. This is a downsampling factor of four so that finally $|D^L_f| = 122$ coarse-resolution grid cells over the UK and Ireland; see Figure \ref{fig:intro}, top panels. Since we were only provided with the interaction function $b_t(\svec,\cdot)$ evaluated with $\svec$ corresponding to the four measurement station locations, we set $D^L_m = D^O_m$.

\subsubsection{Inference}

The final, extant dataset had 3,409 observations for the period January--March 2014. Out of these, we held 20\% for validation purposes and used the other 80\% for model-training purposes. The validation set was chosen by randomly sampling without replacement from all the observations. 

In this application we are interested in computing the total flux by grid cell, in units g s$^{-1}$. This can be carried out directly by adding a grid-area-dependent offset to $\widetilde \mu_f(\svec \mid \varthetab)$ and constructing $\Bmat_{B,t}$ without the grid-area weights (see \ref{sec:correction}). The Laplace-EM algorithm was configured to carry out 10 gradient descents at each M-step, and it converged in 20 iterations to the parameter estimates, $\hat\thetab \equiv (\hat\sigma^2_\zeta,\hat a, \hat d)' =  $(690 ppb$^2,0.972,2.27^\circ)'$. The standard deviation of the discrepancy, $\hat\sigma_\zeta \simeq 26$ ppb,  is considerably larger than that associated with fine-scale temporal variability during each 2 h window and instrumentation error ($\simeq 8.6$ ppb at TAC, for example), and thus it is not negligible. Moreover, the magnitudes of $\hat d = 2.27^\circ$ and $\hat a = 0.972$ are indicative of considerable spatial and temporal dependence.  The parameter $\hat a = 0.972$ corresponds to a temporal-correlation e-folding length of $-2/\ln(\hat a) = 70$ h (where the factor 2 adjusts for the temporal interval used, 2 h).

The case study presented here is illustrative, as it only considers a crude treatment of the background and makes the strong assumption that the fluxes outside the UK and Ireland land areas are known. Yet, it is indicative to compare our results to those in \cite{Ganesan_2015}, where a similar model to ours was used, but without the lognormal spatial assumption. The spatial-correlation e-folding length obtained here ($\hat{d} = 2.27^\circ$) matches within error that obtained in \cite{Ganesan_2015} (posterior median equal to 133 km), but the temporal-correlation e-folding length we obtain is 2--3 times larger than the one in \cite{Ganesan_2015}. Such a difference is possibly because they considered sea areas within the UK and Ireland territory (these were omitted from the model by us), because the result in \cite{Ganesan_2015} is averaged from several monthly studies over a 2-year period, and because we use a lognormal process for the flux field.

\begin{figure}[!t]
\begin{center}
\includegraphics[width=3in]{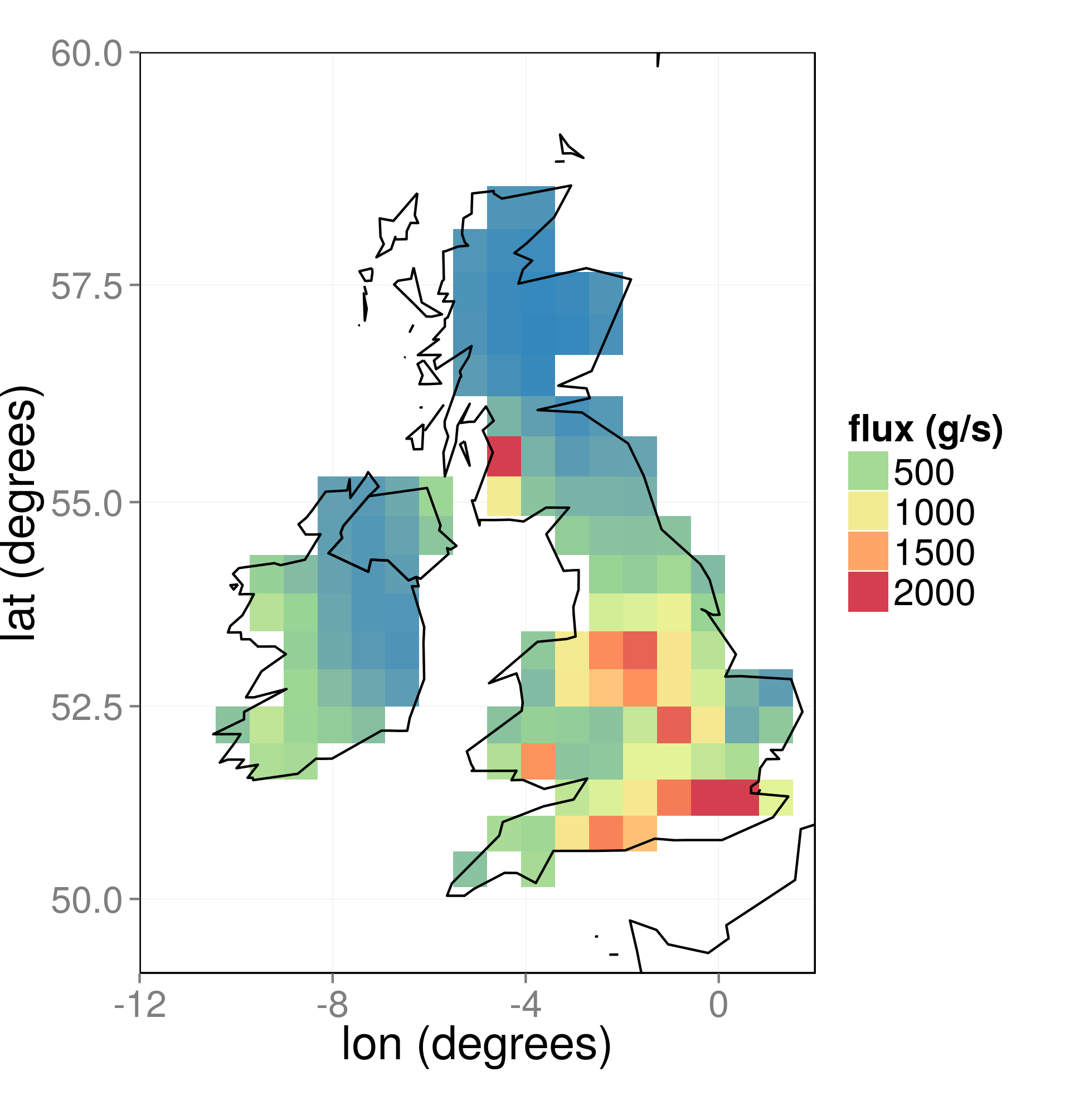}  
\includegraphics[width=3in]{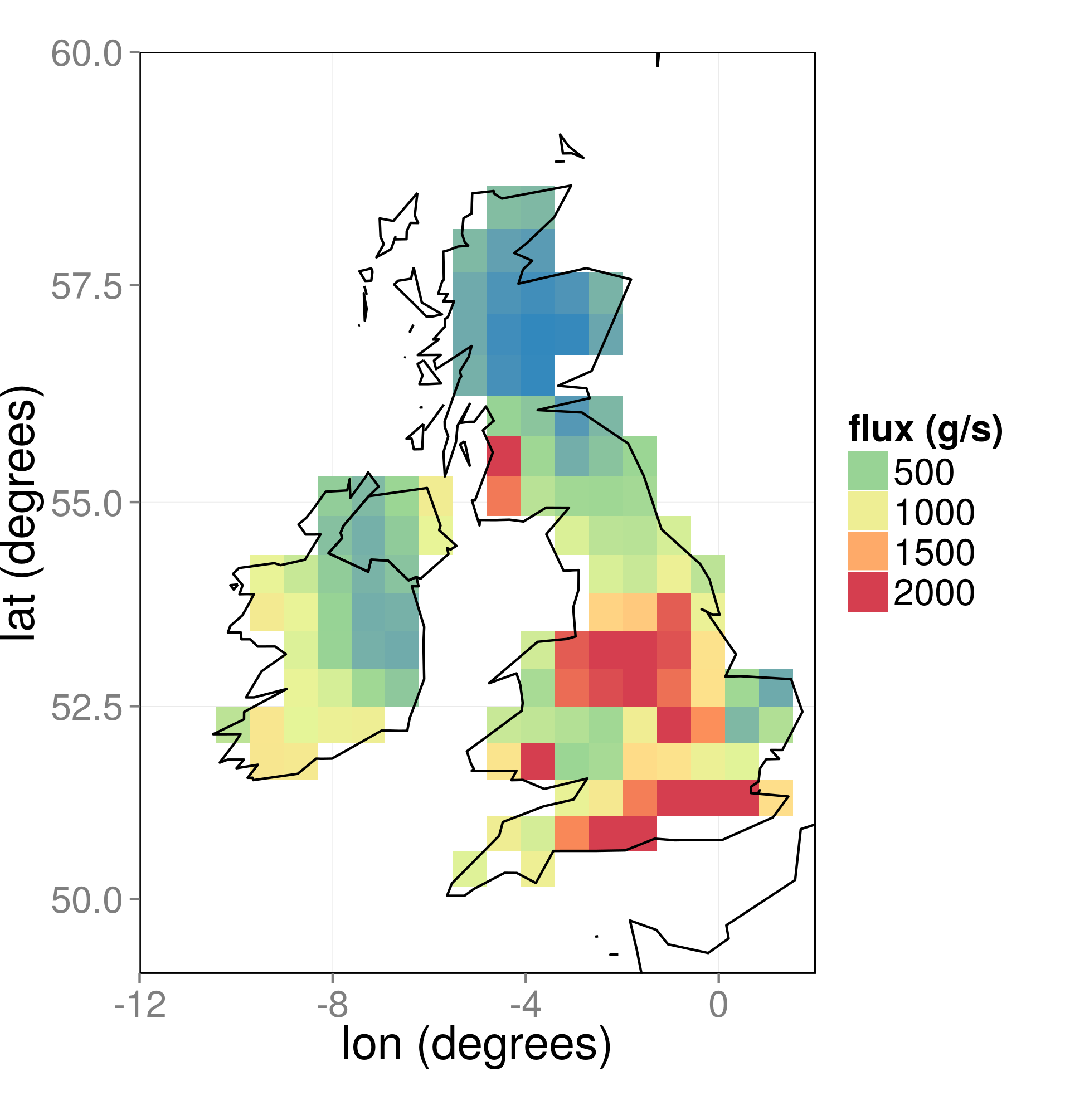}
	\caption{Median (left panel) and 95th percentile (right panel) total methane emissions (fluxes) in the UK and Ireland by grid cell, obtained using our Laplace-EM/HMC approach. White grid cells denote regions were total fluxes were assumed known and used to correct the observation as described in Section \ref{sec:preprocess}.} \label{fig:Em}
\end{center}
\end{figure}

As in the simulation study (Section \ref{sec:sim}), we iterated the HMC sampler 10,000 times in order to obtain samples from the empirical predictive distribution, $[\Yvec_f \mid  \Zvec_m, \hat\thetab, \hat\varthetab]$, with $L = 20$ steps and $\Delta \in [0.070,0.071]$ in the HMC. These settings resulted in an acceptance rate of 54\%. The pointwise posterior median and 95th percentile are shown in the left and right panel, respectively, of Figure \ref{fig:Em}, and they should be compared with the inventory emissions map shown in Figure \ref{fig:intro}, top-left panel. We obtain total estimates for the UK and Ireland (including inventory values for the sea territories) of 2.23 $\pm$ 0.08 Tg yr$^{-1}$ and 0.37 $\pm$ 0.05 Tg yr$^{-1}$, respectively. Both these estimates are lower than those obtained from the inventory shown in Figure \ref{fig:intro}, top-left panel (2.65 Tg yr$^{-1}$ and 0.64 Tg yr$^{-1}$, respectively), and they support the conclusion in  \cite{Ganesan_2015} that these inventory emissions might be too large. 

Interestingly, despite the omission of a spatially varying prior distribution, the posterior spatial distribution of emissions (Figure \ref{fig:Em}) is compatible somewhat with that in the inventory (Figure \ref{fig:intro}, top-left panel). However, there are some key differences. Many large cities, such as Edinburgh, Glasgow, and Dublin, were estimated with lower methane emissions. On the other hand, we obtained larger emissions for the south of the UK, most notably a median total emission of 0.128 Tg yr$^{-1}$ in the grid cell containing London, which compares to 0.062 Tg yr$^{-1}$ in the inventory. We note that there is no consensus on methane emissions in London, but our result supports the observation in \cite{Helfter_2013} that there is a two- to three-fold difference between inventory and measured fluxes in central London.

\begin{figure}[!ht]
\begin{center}
\includegraphics[width=6.3in]{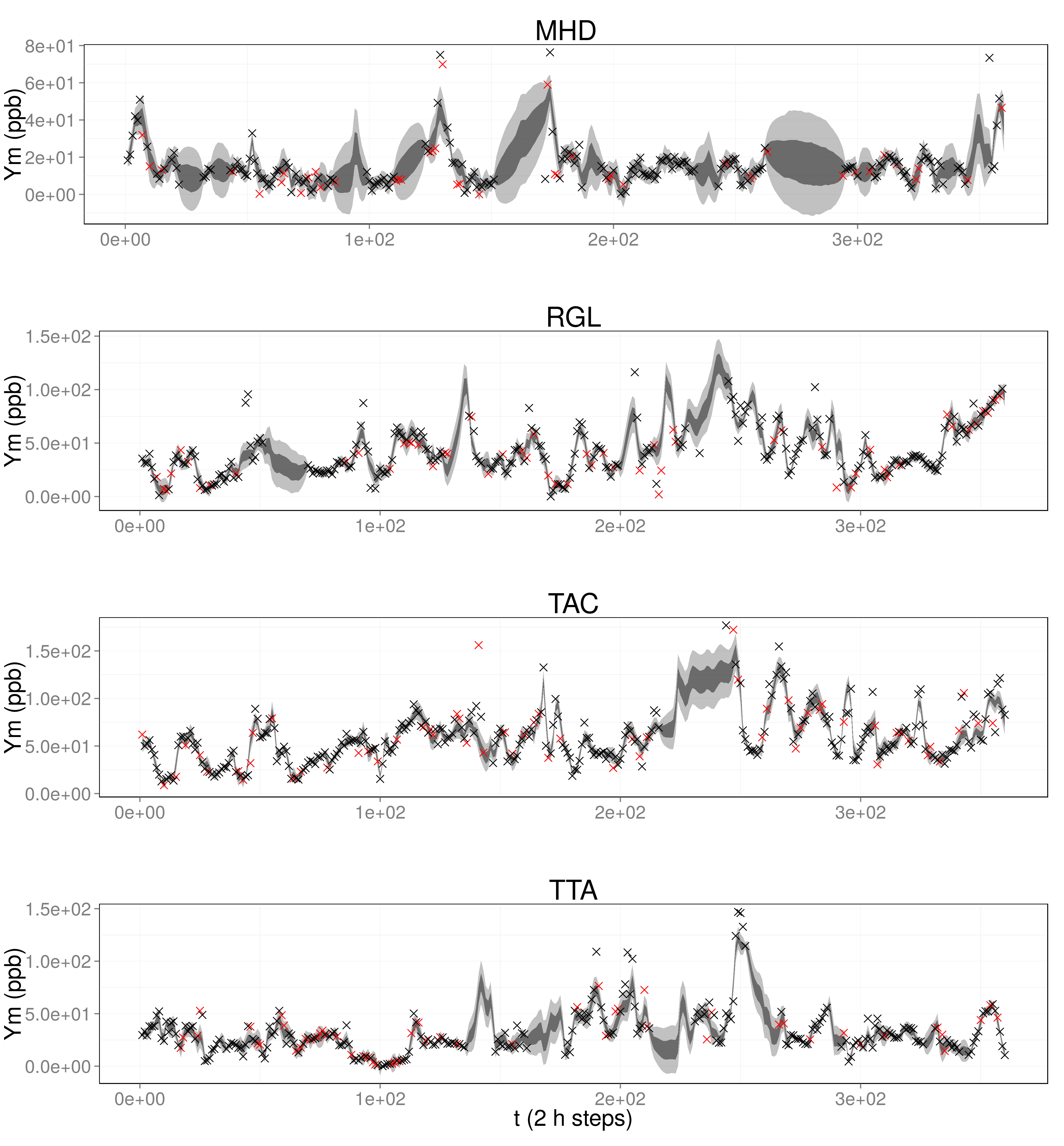}  
	\caption{Distributions of mole fraction (with the background mole fraction subtracted) due to UK and Ireland land-based emissions at the four measurement stations, where $t = 1$ corresponds to 01 January 2014 at 00:00, $t = 2$ corresponds to 01 January 2014 at 02:00, and so on until 31 January 2014 at 22:00. The dark and light shadings denote the interquartile and the 5th--95th percentile ranges, respectively. The red crosses denote the observations used for validation, whilst the black crosses denote those used for training.}\label{fig:MF_est}
\end{center}
\end{figure}

In Figure \ref{fig:MF_est}, we show the distributions of mole fraction at each station location during the month of January, together with both those observations used to train the model (black crosses) and those observations left out for validation (red crosses). The first thing to note is that the distributions of mole fractions are available at times when observations are missing. Mole-fraction uncertainty increases at these times, but clear patterns are also discernible (e.g., at RGL, the peak at $t$ =130 is noticeable). It would be straightforward to show the distribution of mole fractions at other isolated locations, however this requires the interaction function to be evaluated at these locations too. Second, sometimes the posterior density of the mole fraction at a space-time coordinate $(\svec,t)$ might have probability mass on the negative real axis. This does not imply a negative (unphysical) mole fraction, rather an imperfect correction of the background, which in this example is on the order of 1900 ppb.  Third, the observations used for validation lie within the 90\% prediction intervals, with the exception of a few outliers. The ability to obtain what seem to be realistic predictions for out-of-sample mole fractions increases our confidence in inferences for $\Yvec_f$. However, without the availability of validation methane-flux data, critical predictive performance measures in trace-gas inversion cannot be obtained.


\section{Discussion}\label{sec:discussion}

In this article, we present a bivariate spatio-temporal model for flux and mole-fraction fields that allows atmospheric trace-gas inversion. We make use of efficient computational methods, and we include spatial correlations in the flux-field description while maintaining lognormal properties typical of some trace gases. We give a way to estimate these spatial correlations using flux inventories, and we show how to include them within our model. Importantly, uncertainties are captured at each layer in the model (flux level, mole-fraction level, observation level), and predictive distributions of mole fractions include uncertainty on the SRR discrepancy. 

However, as yet we do not capture the uncertainty in estimation of the parameters in the flux and discrepancy models. Consequently,  adjustments might need to be made to the prediction intervals obtained from $[\Yvec_f,\Yvec_m \mid \Zvec_m,\hat\thetab,\hat\varthetab]$ to cater for the uncertainty in $\{\hat\thetab,\hat\varthetab\}$. Of the several approaches that may be used for adjustment, the parametric bootstrap approach of Laird and Louis \cite[][Chapter 3]{Laird_1987,Carlin_2000} appears to be readily applicable here. In the context of this work, one would forward-simulate an ensemble of mole-fraction observations using the parameter estimates $\{\hat\thetab,\hat\varthetab\}$ and from each of these simulated observations, one would re-estimate the parameters using the Laplace-EM algorithm. Then the collection of parameter estimates can be used to obtain a collection of predictive densities (using HMC) that can be averaged to give an adjusted predictive distribution for $\{\Yvec_f,\Yvec_m\}$. Here we expect $\varthetab$ to be unidentifiable, so we suggest bootstrapping  only to adjust for uncertainty in $\thetab$. Alternatively, one might obtain bootstrap estimates of  $\varthetab$ using only the inventory data $\Zmat_{inv}$ (Section \ref{sec:data-feature}).

The lognormal distribution was chosen in order to satisfy the positivity constraint typically imposed on methane flux. As has been noted elsewhere \cite{Miller_2014}, it has the undesirable property that negative fluxes occur with zero probability; negative methane fluxes are unlikely but can occur. This can be remedied by introducing a third, shifting parameter into the lognormal distribution such that $P(Y_f \le 0) > 0$. The resulting distribution, known as the \emph{three-parameter lognormal distribution} in the univariate case \citep[][p.~113]{Johnson_1970}, can also be calibrated using existing inventories, in a manner similar to Section \ref{sec:data-feature}. This increases the model's generality and, of course, distributions other than the lognormal may also be considered.

For computational reasons, we assumed that the discrepancy term has a covariance function that is separable in space and time. However, computational limits will also be reached with this approach, for increasing $|\mathcal{T}|$ or $|D^L_m|$. The cost of storing an $n \times n$ dense matrix is $O(n^2)$, and the computational cost of inverting it is $O(n^3)$. One approach to remedy this is to define sparse inverse covariance matrices for two Gaussian Markov random fields (GMRFs), one for space and one for time, over the space-time locations of the mole-fraction field; see \cite{Rue_2005} for a comprehensive treatment of GMRFs. This sparsity reduces considerably the memory and computational requirements in inference. Space-time separable GMRFs have been used in the context of air-quality monitoring in \cite{Cameletti_2013}. Another approach is through reduced-rank spatio-temporal models \cite{Wikle_2001,Cressie_2010,Katzfuss_2011}, and there spatio-temporal basis functions (but not separability) is required. Other possibilities are through predictive processes \cite{Banerjee_2008} and approximations to stochastic partial differential equations \cite{Lindgren_2011}.

\section*{Acknowledgements}

Noel Cressie's research was partially supported by the 2015--2017 Australian Research Council Discovery Project DP150104576 and partially supported by the National Aeronautics and Space Administration (NASA) grant NNH11ZDA001N-OCO2. Anita Ganesan was funded by the Natural Environment Research Council (NERC) Independent Research Fellowship NE/L010992/1. Measurements at Mace Head, Ridge Hill, Tacolneston, and Angus were funded by the UK Department of Energy and Climate Change (DECC) grant GA0201 to the University of Bristol. Measurements at Mace Head were also partially funded from NASA grant NNX11AF17G to the Massachusetts Institute of Technology, which supports the Advanced Global Atmospheric Gases Experiment (AGAGE) and at Tacolneston through the NERC National Centre for Atmospheric Research. The UK National Atmospheric Emissions Inventory (NAEI) was funded by DECC, the Department for Environment, Food and Rural Affairs (DEFRA), the Scottish Government, the Welsh Government, and the Northern Ireland Department of Environment. We would like to thank Matthew Rigby, Jonathan Rougier, and Dani Gammerman for discussions relevant to this article.



\bibliography{Chemo_bib}

\appendix

\section{The Laplace-EM algorithm for estimating discrepancy parameters}\label{sec:Laplace-EM}

We first describe some of the notation used throughout this appendix. The symbol $\oslash$ will be used to denote element-wise division whilst the symbol $\odot$ denotes element-wise multiplication. These operators can only be used for vectors or matrices of the same size and result in a vector, or matrix, of that size. Throughout, we will use numerator layout notation: The derivative of a scalar by a (column) vector returns a transposed vector, whilst the derivative of a scalar by a matrix $\Xmat$ returns a matrix whose $(i,j)$th element corresponds to the derivative of the scalar with respect to the $(j,i)$th element of $\Xmat$.  

We shall also be repeatedly using the following identities for two vectors $\Xvec, \Yvec$ in the E-step:
\begin{align*}
\diag(\Yvec)\Xvec \equiv  \diag(\Xvec)\Yvec, \\
\frac{\partial\ln \Xvec}{\Xvec} \equiv \diag(\oneb \oslash \Xvec),
\end{align*}
where $\diag(\Xvec)$ returns a diagonal matrix with $\Xvec$ on the diagonal.

In the M-step of the EM algorithm we shall use the following identities for two matrices $\Amat, \Bmat$ of size $p \times p$ and $q \times q$ respectively, and scalar $x$:
\begin{align}
|\Amat^n| &\equiv |\Amat|^n, \nonumber\\
(\Amat \otimes \Bmat)^{-1} &\equiv \Amat^{-1} \otimes \Bmat^{-1}, \nonumber\\
|\Amat \otimes \Bmat | &\equiv |\Amat|^q|\Bmat|^p, \nonumber\\
\frac{\partial \ln |\Amat|}{\partial x} &\equiv \trace\left(\Amat^{-1}\frac{\partial\Amat}{\partial x}\right), \label{eq:deriv1}\\
\frac{\partial(\Amat \otimes \Bmat)}{\partial x} &\equiv \Amat \otimes \frac{\partial \Bmat}{\partial x} + \frac{\partial \Amat}{\partial x} \otimes \Bmat, \label{eq:deriv2}
\end{align}
where in this context $|\,\cdot\,|$ is the determinant and $\otimes$ is the Kronecker product. 

\subsection{E-Step}\label{sec:Estep}

In the E-step, we compute 
\begin{equation}\label{eq:Q}
Q(\thetab\mid  \thetab^{(i)}) \equiv \E(\ln L_c(\thetab) \mid \Zvec_m , \thetab^{(i)},\hat\varthetab),
\end{equation}
that is, the expectation of the complete-data log-likelihood with respect to the conditional distribution of $\{\Yvec_f,\Yvec_m\}$ given the data $\Zvec_m$, $\hat\varthetab$, and the current parameter estimate $\thetab^{(i)}$, $i = 0,1,\dots$. Hence, in order to be able to evaluate $Q(\thetab\mid  \thetab^{(i)})$, we first need to determine the conditional distribution under which to take expectations. 

Define $\Qmat_O \equiv \sigma^{-2}_\varepsilon \Imat$, $\Qmat_\zeta(\thetab^{(i)}) \equiv \Sigmamat_\zeta^{-1}(\thetab^{(i)})$, and $\Qmatt_f(\hat\varthetab) = \Sigmamatt_f^{-1}(\hat\varthetab)$, where
\begin{align*}
\Sigmamat_\zeta(\thetab^{(i)}) &\equiv (\cov(\zetab_t(\svec),\zetab_{t'}(\uvec)\mid  \thetab^{(i)}): \svec, \uvec \in D^L_m;\, t,t' \in \mathcal{T}), \\
\Sigmamatt_f(\hat\varthetab) &\equiv (\widetilde{C}_\ff(\svec,\uvec\mid  \hat\varthetab): \svec, \uvec \in D^L_f).
\end{align*} 

\noindent For notational convenience, from now on we shall omit the dependence of $\Sigmamatt_f$ and $\Qmatt_f$ on $\hat\varthetab$, and the dependence of $\Sigmamat_\zeta$ and $\Qmat_\zeta$ on $\thetab^{(i)}$.

The observation model for the discretised system is,
\begin{equation*}
\Zvec_m = \Cmat \Yvec_m + \epsilonb_m,
\end{equation*}
where $\Cmat \equiv (\Cmat'_t : t\in\mathcal{T})'$ is the incidence matrix mapping the discretised mole-fraction field to the observations. Define $\Bmat_B \equiv(\Bmat_{B,t}' : t \in \mathcal{T})'$ and $\muvect_f \equiv (\widetilde{\mu}_f(\svec \mid \hat\varthetab) : \svec \in D^L_f)'$. Then 
\begin{align}\label{eq:log-lik}
\ln [\Yvec_f, \Yvec_m \mid  \Zvec_m, \thetab^{(i)},\hat\varthetab] = c_1 - &0.5 (\Zvec_m - \Cmat\Yvec_m)'\Qmat_O(\Zvec_m - \Cmat\Yvec_m) \nonumber \\
& - 0.5(\Yvec_m - \Bmat_B\Yvec_f)'\Qmat_\zeta(\Yvec_m - \Bmat_B\Yvec_f) \\
& - 0.5(\ln\Yvec_f - \muvect_f)'\Qmatt_f(\ln \Yvec_f - \muvect_f) - (\ln\Yvec_f)'\oneb. \nonumber
\end{align}
Due to the presence of $\ln \Yvec_f$ in \eqref{eq:log-lik}, this conditional distribution is not Gaussian. We summarise it using the first two moments, approximated using first-order and second-order derivatives.

Define $\Dmat_f \equiv \diag(\oneb \oslash \Yvec_f)$ and $\Dmat_\ff \equiv \diag(\oneb \oslash (\Yvec_f \odot \Yvec_f))$. Then the derivatives $\Jmat$ of \eqref{eq:log-lik} with respect to $\Yvec \equiv (\Yvec_f', \Yvec_m')'$, are given by

\begin{align*}
\Jmat(\Yvec\mid  \thetab^{(i)},\hat\varthetab) &\equiv \begin{pmatrix} \displaystyle \frac{\partial \ln [\Yvec_f, \Yvec_m \mid  \Zvec_m, \thetab^{(i)},\hat\varthetab]} {\partial \Yvec_f' } \vspace{0.1in}\\
                        \displaystyle \frac{\partial \ln [\Yvec_f, \Yvec_m \mid  \Zvec_m, \thetab^{(i)},\hat\varthetab]} {\partial \Yvec_m'} \end{pmatrix}  \\ 
&= \begin{pmatrix} \Bmat_B'\Qmat_\zeta\Yvec_m - \Bmat_B'\Qmat_\zeta\Bmat\Yvec_f - \Dmat_f\Qmatt_f(\ln\Yvec_f - \muvect_f) - (\oneb \oslash \Yvec_f) \\ -\Cmat'\Qmat_O\Cmat\Yvec_m + \Cmat'\Qmat_O\Zvec_m - \Qmat_\zeta(\Yvec_m -\Bmat_B\Yvec_f)\end{pmatrix}.
\end{align*}

The Hessian $\Hmat$ is

\begin{align*}
\Hmat(\Yvec\mid \thetab^{(i)},\hat\varthetab)  &\equiv \begin{pmatrix} \displaystyle \frac{\partial \ln [\Yvec_f, \Yvec_m \mid  \Zvec_m, \thetab^{(i)},\hat\varthetab]} {\partial \Yvec_f\Yvec_f' } &
                        \displaystyle \frac{\partial \ln [\Yvec_f, \Yvec_m \mid  \Zvec_m, \thetab^{(i)},\hat\varthetab]} {\partial \Yvec_m\Yvec_f' }\vspace{0.1in} \\
                        \displaystyle \frac{\partial \ln [\Yvec_f, \Yvec_m \mid  \Zvec_m, \thetab^{(i)},\hat\varthetab]} {\partial \Yvec_f\Yvec_m' } &
                        \displaystyle \frac{\partial \ln [\Yvec_f, \Yvec_m \mid  \Zvec_m, \thetab^{(i)},\hat\varthetab]} {\partial \Yvec_m\Yvec_m'} \end{pmatrix} \nonumber \\
      &= \begin{pmatrix} -\Bmat_B'\Qmat_\zeta\Bmat_B - \Dmat_f\Qmatt_f\Dmat_f + \diag(\Qmatt_f (\ln \Yvec_f - \muvect_f)) \Dmat_{\ff} +\Dmat_{\ff}& \Bmat_B'\Qmat_\zeta \\ \Qmat_\zeta\Bmat_B & -\Cmat' \Qmat_O \Cmat - \Qmat_\zeta\end{pmatrix}. 
\end{align*}
We now use a gradient-descent algorithm with the derivatives $\Jmat(\Yvec \mid \thetab^{(i)},\hat\varthetab)$ to find the mode $\Yvec^*$ of the distribution and use this as an approximation to the mean. We evaluate $\Hmat$ at this mode; the approximate covariance is then given by $-\Hmat(\Yvec^* \mid \thetab^{(i)},\hat\varthetab)^{-1}$. For gradient descent, we used the $R$ function \texttt{optim} with the method \texttt{BFGS}.

From \eqref{eq:Q}, the function $Q(\thetab \mid \thetab^{(i)})$ is
\begin{align*}
Q(\thetab \mid \thetab^{(i)}) &= -0.5 \ln |\Sigmamat_\zeta(\thetab)| - 0.5\trace(\Qmat_\zeta(\thetab)\Psib), \nonumber
\end{align*}
\noindent where 
\begin{align} 
\Psib \equiv \, &\E(\Yvec_m \Yvec_m'\mid \Zvec_m , \thetab^{(i)},\hat\varthetab) \nonumber \\
             &+ \Bmat_B \E(\Yvec_f \Yvec_f'\mid \Zvec_m , \thetab^{(i)},\hat\varthetab) \Bmat_B' \label{eq:Psi} \\
             &- 2\Bmat_B \E( \Yvec_f \Yvec_m'\mid \Zvec_m , \thetab^{(i)},\hat\varthetab). \nonumber
\end{align}\label{sec:Psisec}
Note that only the first two moments of $(\Yvec_f',\Yvec_m')'$ are required to compute \eqref{eq:Psi}.

\subsection{M-step}\label{sec:Mstep}

In the M-step we set

\begin{equation}
\thetab^{(i+1)} = \argmax_{\thetab} Q(\thetab\mid \thetab^{(i)}).\label{eq:thetamax}
\end{equation}
The discrepancy term $\zeta_t(\svec)$ has a separable spatio-temporal covariance function. Hence, we can write the full space-time covariance matrix as 
\begin{align}
\Sigmamat_\zeta(\thetab) &= \sigma^2_\zeta\Rmat_{\zeta}(a,d) \nonumber \\
                         &= \sigma^2_\zeta\Rmat_{\zeta,t}(a) \otimes \Rmat_{\zeta,s}(d), \label{eq:Kron_form}
\end{align}
where recall that $\otimes$ is the Kronecker product and $\Rmat_{\zeta,s}$, $\Rmat_{\zeta,t}$ are correlation matrices. In the M-step we maximise $Q(\thetab\mid \thetab^{(i)})$ in \eqref{eq:thetamax} using gradient descent; the following gradients are obtained by repeated application of \eqref{eq:deriv1} and \eqref{eq:deriv2}:

\begin{align*}
\frac{\partial Q(\thetab\mid \thetab^{(i)})}{\partial \sigma_\zeta^{2}} &\equiv 0.5\left(-n\sigma_\zeta^{-2} + \sigma_\zeta^{-4}\trace(\Rmat_\zeta(a,d)^{-1}\Psib)\right), \\ & \\
\frac{\partial Q(\thetab\mid \thetab^{(i)})}{\partial a} &\equiv -0.5 \trace\left((\Rmat_{\zeta,t}(a)\otimes\Rmat_{\zeta,s}(d))^{-1}\left(\frac{\partial \Rmat_{\zeta,t}(a)}{\partial a}  \otimes \Rmat_{\zeta,s}(d)\right)\right)\\  
&~~+ 0.5 \sigma_\zeta^{-2}\trace\left(\Rmat_\zeta^{-1}(a,d)\left(\frac{\partial \Rmat_{\zeta,t}(a)}{\partial a} \otimes \Rmat_{\zeta,s}(d)\right) \Rmat_\zeta^{-1}(a,d)\Psib \right), \\ & \\
\frac{\partial Q(\thetab\mid \thetab^{(i)})}{\partial d} &\equiv -0.5 \trace\left((\Rmat_{\zeta,t}(a)\otimes\Rmat_{\zeta,s}(d))^{-1}\left(\Rmat_{\zeta,t}(a) \otimes\frac{\partial \Rmat_{\zeta,s}(d)}{\partial d}\right)\right) \label{eq:grad_d}\\ 
&~~+ 0.5 \sigma_\zeta^{-2}\trace\left(\Rmat_\zeta^{-1}(a,d)\left(\Rmat_{\zeta,t}(a) \otimes \frac{\partial \Rmat_{\zeta,s}(d)}{\partial d}\right) \Rmat_\zeta^{-1}(a,d)\Psib \right),
\end{align*}
where here $n = |D^L_m||\mathcal{T}|$. In the equations above, the mixed-product property of the Kronecker product can be used to simplify the computations; for example,
 \begin{equation*}
 (\Rmat_{\zeta,t}(a)\otimes\Rmat_{\zeta,s}(d))^{-1}\left(\frac{\partial \Rmat_{\zeta,t}(a)}{\partial a}  \otimes \Rmat_{\zeta,s}(d)\right)  \equiv \left(\Rmat^{-1}_{\zeta,t}(a)\frac{\partial \Rmat_{\zeta,t}(a)}{\partial a}\right)\otimes \Imat.
 \end{equation*}

All that remains is to specify the partial derivatives of the spatial and temporal correlation matrices with respect to $a$ and $d$. These are obtained from the derivatives of the correlation functions:
\begin{align*}
\frac{\partial}{\partial a}\rho_t(t,t') &\equiv |t - t'|a^{|t -t'| - 1}; \quad |a| < 1, \\
\frac{\partial}{\partial d}\rho_s(\svec,\uvec) &\equiv \frac{\|\uvec - \svec \|}{d^2}\exp\left(-\frac{\| \uvec - \svec \|}{d} \right); \quad d > 0.
\end{align*}

\section{Directly inferring the total flux by grid cell}\label{sec:correction}

Instead of computing the flux (a density that is per unit area), running the EM-Laplace/HMC algorithm, and rescaling again to obtain the total flux, we can instead infer the total flux by grid cell, $\{ A(\uvec) Y_f(\uvec): \uvec \in D^L_f\}$, directly with only a few modifications. Write down the approximation to the integral in \eqref{eq:mv} as
\begin{align*}
\int_D b_t(\svec,\uvec)Y_f(\uvec)\intd\u &\simeq \sum_{\uvec \in D^L_f}b_t(\svec,\uvec)(A(\uvec)Y_f(\uvec))\\
& = \sum_{\uvec \in D^L_f}b_t(\svec,\uvec)Y_f^{tot}(\uvec),
\end{align*}
where $Y_f^{tot}(\uvec) \equiv A(\uvec)Y_f(\uvec)$. Now we seek a model for $Y_f^{tot}(\uvec)$ ; however, this poses no additional work since 
\begin{equation}\label{eq:log-shift}
\widetilde{Y}_f^{tot}(\svec) \equiv \ln Y_f^{tot}(\svec) = \ln A(\svec) + \ln Y_f(\svec).
\end{equation}
It thus follows that  $Y_f^{tot}$ is also a lognormal spatial process, shifted by a known quantity. From \eqref{eq:log-shift}, 
\begin{align*}
\E(\widetilde{Y}_f^{tot}(\svec)\mid \varthetab) &= \ln A(\svec) + \widetilde{\mu}_f(\svec \mid \varthetab), \\
\cov(\widetilde{Y}_f^{tot}(\svec),\widetilde{Y}_f^{tot}(\uvec) \mid \varthetab) &= \widetilde{C}_{\ff}(\s,\u \mid \varthetab).
\end{align*}
Note that the covariance function remains unchanged, and in order to solve directly for $\widetilde{Y}_f^{tot}(\cdot)$ one uses $\widetilde{\mu}_f^{tot}(\svec) \equiv \ln A(\svec) + \widetilde{\mu}_f(\svec \mid \varthetab)$, instead of $\widetilde{\mu}_f(\svec \mid \varthetab)$ in \eqref{eq:ln1} and \eqref{eq:ln2}. Then $\Bmat_{B,t}$ is constructed \emph{without} the area-weight terms since they have already been accounted for. That is, in this case the matrix $\Bmat_{B,t}$ is given by,
\begin{equation*}
\Bmat_{B,t} \equiv (b_t(\svec,\uvec): \svec \in D^L_m;\,	 \uvec \in D^L_f).
\end{equation*}

\end{document}